\documentclass[12pt,eqno,epsf]{article}
\usepackage{amsmath,amssymb,graphicx,slashbox}
\numberwithin{equation}{section}

\def\ignore#1{{}}

\newcounter{sxn}

\newcounter{axn}

\date{}

\newdimen\mybaselineskip
\renewcommand{\thefootnote}{\arabic{footnote}}

\newcommand{\beeq}{\begin{equation}}
\newcommand{\eneq}{\end{equation}}
\newcommand{\beqn}{\begin{eqnarray}}
\newcommand{\eeqn}{\end{eqnarray}}

\newcommand{\alp}{\alpha}
\newcommand{\bt}{\beta}
\newcommand{\gm}{\gamma}
\newcommand{\Gm}{\Gamma}
\newcommand{\dlt}{\delta}

\newcommand{\ep}{\epsilon}
\newcommand{\tht}{\theta}

\newcommand{\Tht}{\Theta}

\newcommand{\kp}{\kappa}
\newcommand{\lmd}{\lambda}
\newcommand{\Lmd}{\Lambda}
\newcommand{\sgm}{\sigma}
\newcommand{\Sgm}{\Sigma}
\newcommand{\Ups}{\Upsilon}

\newcommand{\dalp}{\dot{\alpha}}
\newcommand{\dbt}{\dot{\beta}}
\newcommand{\dgm}{\dot{\gamma}}
\newcommand{\ddlt}{\dot{\delta}}

\newcommand{\ualp}{\underline{\alp}}
\newcommand{\ubt}{\underline{\bt}}
\newcommand{\ugm}{\underline{\gm}}
\newcommand{\udlt}{\underline{\dlt}}

\newcommand{\be}{\begin{equation}}
\newcommand{\ee}{\end{equation}}
\newcommand{\bea}{\begin{eqnarray}}
\newcommand{\eea}{\end{eqnarray}}
\newcommand{\eql}{\!\!\!&=\!\!\!&}

\newcommand{\defa}{\!\!\!&\equiv\!\!\!&}

\newcommand{\tl}[1]{\tilde{#1}}
\newcommand{\bdm}[1]{{\mbox{\boldmath $#1$}}}
\newcommand{\sbdm}[1]{\mbox{\scriptsize \boldmath $#1$}}

\newcommand{\diag}{{\rm diag}}
\newcommand{\der}{\partial}
\newcommand{\dr}{\!\!d}
\newcommand{\hc}{{\rm h.c.}}
\newcommand{\ie}{{i.e.}}
\newcommand{\id}{\mbox{\boldmath $1$}}

\newcommand{\brkt}[1]{\left( #1 \right)}
\newcommand{\brc}[1]{\left\{ #1 \right\}}
\newcommand{\sbk}[1]{\left[ #1 \right]}
\newcommand{\abs}[1]{\left| #1 \right|}

\renewcommand{\Im}{{\rm Im}\,}

\newcommand{\cA}{{\cal A}}

\newcommand{\cD}{{\cal D}}

\newcommand{\cF}{{\cal F}}

\newcommand{\cL}{{\cal L}}

\newcommand{\cT}{{\cal T}}
\newcommand{\cV}{{\cal V}}

\newcommand{\cW}{{\cal W}}

\newcommand{\cPT}{{\cal P}_{\rm T}}

\begin{document}
\thispagestyle{empty}

\baselineskip=12pt

%{\small \noindent \mydate \hfill }

\begin{flushright}
KEK-TH-1792 \\
WU-HEP-15-02
\end{flushright}

\baselineskip=35pt plus 1pt minus 1pt

\vskip 1.5cm

\begin{center}
{\LARGE\bf $\bdm{N=1}$ superfield description of \\
vector-tensor couplings 
in six dimensions}

\vspace{1.5cm}
\baselineskip=20pt plus 1pt minus 1pt

\normalsize

{\large\bf Hiroyuki Abe,}${}^1\!${\def\thefootnote{\fnsymbol{footnote}}
\footnote[1]{E-mail address: abe@waseda.jp}}
{\large\bf Yutaka Sakamura}${}^{2,3}\!${\def\thefootnote{\fnsymbol{footnote}}
\footnote[2]{E-mail address: sakamura@post.kek.jp}} 
{\large\bf and Yusuke Yamada}${}^1\!${\def\thefootnote{\fnsymbol{footnote}}
\footnote[3]{E-mail address: yuusuke-yamada@asagi.waseda.jp}}

\vskip 1.0em

${}^1${\small\it Department of Physics, Waseda University, \\ 
Tokyo 169-8555, Japan}

\vskip 1.0em

${}^2${\small\it KEK Theory Center, Institute of Particle and Nuclear Studies, 
KEK, \\ Tsukuba, Ibaraki 305-0801, Japan} \\ \vspace{1mm}
${}^3${\small\it Department of Particles and Nuclear Physics, \\
The Graduate University for Advanced Studies (Sokendai), \\
Tsukuba, Ibaraki 305-0801, Japan}

\end{center}

\vskip 1.0cm
\baselineskip=20pt plus 1pt minus 1pt

\begin{abstract}
We express supersymmetric couplings 
among the vector and the tensor multiplets in six dimensions (6D) 
in terms of $N=1$ superfields. 
The superfield description is derived from 
the invariant action in the projective superspace. 
The obtained expression is consistent 
with the known superfield actions  
of 6D supersymmetric gauge theory and 5D Chern-Simons theory 
after the dimensional reduction. 
Our result provides a crutial clue to the $N=1$ superfield description 
of 6D supergravity. 
\end{abstract}

%%%%%%%%%%%%%%%%%%%%%%%%
\newpage

\section{Introduction}
The $N=1$ superfield description\footnote{ 
``$N=1$'' denotes supersymmetry with four supercharges in this paper. 
}
of higher dimensional 
supersymmetric (SUSY) theories is quite useful 
when we discuss phenomenological SUSY models 
with extra dimensions. 
It makes the derivation of 4-dimensional (4D) effective theories easier 
since the Kaluza-Klein mode expansion 
can be performed keeping the $N=1$ off-shell structure. 
Besides, the action is expressed compactly, 
and general setups can be treated. 
Since higher-dimensional SUSY theories have extended SUSY, 
the full off-shell formulations are complicated and less familiar, 
or do not even exist 
for theories higher than 6 dimensions (6D). 
In contrast, the $N=1$ superfield description is always possible 
because it only respects part of the full off-shell SUSY structure. 
Hence it is powerful especially when we describe 
interactions between sectors whose dimensions are different, 
such as the bulk-boundary couplings in 5D theories compactified on $S^1/Z_2$. 
For the above reasons, a lot of works along this direction 
have been published~\cite{ArkaniHamed:2001tb}-\cite{Sakamura:2012bj}. 
%Especially, Ref.~\cite{ArkaniHamed:2001tb} provides 
%the $N=1$ description of global SUSY theories 
%in 5-10 dimensions. 

When we discuss a realistic extra-dimensional models, 
the moduli play important roles. 
They have to be stabilized to finite values 
by some mechanism, and are often relevant to the mediation of SUSY-breaking 
to the visible sector. 
In order to treat the moduli properly, 
we need to consider supergravity (SUGRA). 
The $N=1$ superfield description of 5-dimensional (5D) SUGRA is already obtained  
in Refs.~\cite{Linch:2002wg}-%,Paccetti:2004ri,Abe:2004ar,Kuzenko:2005sz,
\cite{Sakamura:2012bj}. 
Making use of it, the moduli dependence of the 4D effective action can 
systematically be derived~\cite{Luty:1999cz}-\cite{Abe:2011rg}.
Our aim is to extend the 5D superfield action to 6D. 
%construct the $N=1$ description of 6-dimensional (6D) SUGRA. 
Since the minimal number of SUSY is the same 
in the 5D and 6D cases, the desired $N=1$ description is expected to be similar 
to that of 5D theories. 
However, there is an obstacle to a straightforward extension 
of the 5D result. 
%First, although Refs.~\cite{Paccetti:2004ri,Abe:2004ar} are based on 
%the superconformal formulation of 5D SUGRA~\cite{Kugo:2000hn}-\cite{Kugo:2002js}, 
%there are no full superconformal off-shell description 
%of the hypermultiplet in 6D case~\cite{Kugo:2000hn}. 
In contrast to the 5D case, 
the 6D superconformal Weyl multiplet contains 
an anti-self-dual antisymmetric tensor~$T^-_{MNL}$ 
($M,N,L$: 6D Lorentz indices)~\cite{Bergshoeff:1985mz}. 
This leads to a difficulty for the Lagrangian formulation, 
similar to that for type IIB SUGRA. 
This difficulty can be evaded by introducing a tensor multiplet, 
which contains an antisymmetric tensor~$B^+_{MN}$ 
whose field strength~$F^+_{MNL}\equiv \der_{[M}B^+_{NL]}$ 
is subject to the self-dual constraint~\cite{Bergshoeff:1985mz}. 
Combining this multiplet with the Weyl multiplet, 
we obtain a new multiplet\footnote{
This is called the ``Weyl 2 multiplet'' in Ref.~\cite{Coomans:2011ih}, 
and the ``type-II Weyl multiplet'' in Ref.~\cite{Linch:2012zh}. 
} 
that contains an unconstrained antisymmetric tensor~$B_{MN}$, 
whose field strength is given by the sum of $T^-_{MNL}$ and $F^+_{MNL}$.  
Namely, the off-shell formulation of 6D SUGRA requires 
the existence of the tensor field~$B_{MN}$, which is not 
a necessary ingredient in 5D SUGRA.\footnote{
Note also that an antisymmetric rank-2 tensor field is dual to a vector field 
in 5 dimensions.
}

The off-shell action of 6D SUGRA is provided 
in Refs.~\cite{Bergshoeff:1985mz,Coomans:2011ih}. 
In that action, the tensor field~$B_{MN}$ is coupled to the vector fields. 
Thus, in this paper, we clarify how the vector-tensor couplings 
are expressed in terms of $N=1$ superfields. 
Since we focus on these couplings, we do not consider 
the gravitational couplings in this paper. 
In this sense, this work is the generalization of 
Ref.~\cite{ArkaniHamed:2001tb} including the vector-tensor couplings. 
For our purpose, the projective superspace 
formulation~\cite{Karlhede:1984vr,Lindstrom:1987ks,Lindstrom:1989ne}
is useful. 
In fact, the $N=1$ superfield description of 5D SUGRA 
can be derived from the action in the projective superspace~\cite{Kuzenko:2005sz}. 
As for 6D SUGRA, the off-shell action in this formulation is provided 
in Ref.~\cite{Linch:2012zh}.\footnote{
The 6D action in the harmonic superspace is provided in Ref.~\cite{Buchbinder:2014sna}. 
}
We derive the $N=1$ superfield action from it. 

The paper is organized as follows. 
In Sec.~\ref{prj_ssp}, we provide a brief review of 
$N=2$ supersymmetric actions in the projective superspace. 
In Sec.~\ref{N1sf_action}, we decompose $N=2$ superfields 
into $N=1$ superfields, and express the vector-tensor couplings 
in terms of the latter. 
We also clarify the relation between our result 
and the known $N=1$ superfield description of 
6D SUSY gauge theory or 5D SUSY Chern-Simons theory 
through the dimensional reduction. 
Sec.~\ref{summary} is devoted to the summary. 
In the appendices, we list our notations for spinors, 
and show explicit derivations of some of the results in the text.

\section{Invariant action in projective superspace} \label{prj_ssp}
\subsection{Action formula}
An $N=2$ off-shell action can be constructed 
by using the projective superspace 
formulation~\cite{Karlhede:1984vr,Lindstrom:1987ks,Lindstrom:1989ne}. 
We consider 6D $(1,0)$ SUSY theories. 
The 6D projective superspace is parametrized by 
the spacetime coodinates~$x^M$ ($M=0,1,\cdots,5$), 
the Grassmannian coordinates~$\Tht_{\ualp}^i$ ($i=1,2$; $\ualp=1,2,3,4$),\footnote{
In this paper, $\underline{\alp},\underline{\bt},\cdots$ denote 
the 4-component spinor indices, 
and $\alp,\bt,\cdots$ and $\dot{\alp},\dot{\bt},\cdots$ are used as 
the 2-component indices 
of 4D SL(2,${\mathbb C}$) spinors. 
} 
which form an SU(2)-Majorana-Weyl spinor, 
and the complex coordinate~$\zeta$ of $\mathbb{CP}^1$. 
A projective superfield~$\Xi(x,\Tht,\zeta)$ is a holomorphic function in $\zeta$ 
that satisfies~\footnote{
The index~$[k]$ indicates the weight-$k$ quantity. 
It coincides with the superconformal weight 
in the superconformal theories~\cite{Kuzenko:2010bd}. 
} 
\be
 \cD_{\ualp}^{[1]}\Xi \equiv \brkt{-\zeta\cD^1_{\ualp}+\cD^2_{\ualp}}\Xi = 0, 
 \label{cstrt:Xi}
\ee
where the spinor derivatives are defined in (\ref{def:cD}). 
It can be expanded as 
\be
 \Xi(x,\Tht,\zeta) = \sum_{n=-\infty}^\infty \Xi_n(x,\Tht)\zeta^n, 
 \label{expand:Xi}
\ee
where $N=2$ superfields~$\Xi_n$ satisfy
\be
 \cD^1_{\ualp}\Xi_n = \cD^2_{\ualp}\Xi_{n+1}. \label{cstrt_Xi}
\ee
The constraint~(\ref{cstrt_Xi}) fixes the dependence of $\Xi_n$ 
on half of the Grassmann coordinates~$\Tht_{\ualp}^i$, and thus 
$\Xi_n$ can be considered as superfields which effectively live 
on an $N=1$ superspace. 

The natural conjugate operation in the projective superspace is 
the combination of the complex conjugate and the antipodal map on $\mathbb{CP}^1$ 
($\zeta^*\to -1/\zeta$), which is called the smile conjugate denoted as
\be
 \breve{\Xi}(x,\Tht,\zeta) = 
 \sum_{n=-\infty}^\infty (-1)^n\bar{\Xi}_{-n}(x,\Tht)\zeta^n. 
\ee
Then the $N=2$ SUSY invariant action formula is given by~\cite{Linch:2012zh,Gates:2005mc}
\be
 S = \int\dr^6x\brc{\left.
 \oint_C\frac{d\zeta}{2\pi i}\;\zeta\cD^{[-4]}L(x,\Tht,\zeta)\right|_{\Tht=0}}, 
 \label{action_fml}
\ee
where $C$ is a contour surrounding the origin~$\zeta=0$, 
the ``Lagrangian superfield''~$L(x,\Tht,\zeta)$ is 
a smile-real projective superfield ($\breve{L}=L$), 
and 
\bea
 \cD^{[-4]} \defa -\frac{1}{96}\ep^{\ualp\ubt\ugm\udlt}
 \cD_{\ualp}^{[-1]}\cD_{\ubt}^{[-1]}\cD_{\ugm}^{[-1]}\cD_{\udlt}^{[-1]}, \nonumber\\
 \cD_{\ualp}^{[-1]} \defa \frac{1}{1+\zeta\eta}\brkt{\cD^1_{\ualp}+\eta\cD^2_{\ubt}}. 
\eea
The complex number~$\eta$ is chosen arbitrarily 
as long as $1+\zeta\eta\neq 0$. 
In fact, the action~(\ref{action_fml}) is independent of $\eta$.

\subsection{Explicit forms of Lagrangians}  \label{VTsector}
A 6D hypermultiplet is described by an arctic superfield~$\Ups$, 
which is a projective superfield that is non-singular 
at the north pole of $\mathbb{CP}^1$ ($\zeta=0$). 
Namely, it is expanded as
\be
 \Ups(x,\Tht,\zeta) = \sum_{n=0}^\infty\Ups_n(x,\Tht)\zeta^n. 
 \label{expand:Ups}
\ee
A 6D vector multiplet is described by a tropical superfield~$\bdm{V}$, 
which is a smile-real projective superfield, 
\be
 \breve{\bdm{V}}(x,\Tht,\zeta) = \bdm{V}(x,\Tht,\zeta). 
\ee
Namely, it is expanded as 
\be
 \bdm{V}(x,\Tht,\zeta) = \sum_{n=-\infty}^\infty V_n(x,\Tht)\zeta^n, \;\;\;\;\;
 V_{-n} = (-1)^n\bar{V}_n. \label{tropical}
\ee
Using these projective superfields, the Lagrangian superfield~$L$ 
in the hypermultiplet sector is given by
\be
 L_{\rm hyper} = \breve{\Ups}e^{-\sbdm{V}}\Ups.  
 \label{L_hyp}
\ee

In the following, we consider Abelian gauge theories, for simplicity. 

In contrast to the above multiplets, 
a 6D tensor multiplet is not described by a projective superfield. 
As first shown in Ref.~\cite{Sokatchev:1988aa}, 
it can be described by a constrained real superfield~$\Phi$ that satisfies 
\be
 \cD_{\ualp}^{(i}\cD_{\ubt}^{j)}\Phi = 0.  \label{cstrt:tsrI}
\ee
or equivalently described by an SU(2)-Majorana-Weyl spinor 
superfield~$T^{i\ualp}$ constrained by 
\be
 \cD_{\ualp}^{(i}T^{j)\ubt}-\frac{1}{4}\dlt_{\ualp}^{\;\;\ubt}
 \cD_{\ugm}^{(i}T^{j)\ugm} = 0. \label{cstrt:tsrII}
\ee
where the parentheses denote the symmetrization for the indices. 
We can identify these superfields as 
\be
 \Phi = \cD_{i\ualp}T^{i\ualp} 
 = \ep_{ij}\cD^j_{\ualp}T^{i\ualp}, \label{rel:Phi-T} 
\ee
but we can also regard them as independent tensor multiplets. 
From these two superfields, we can construct a projective composite superfield, 
\be
 \cT^{[2]} \equiv \frac{i}{\zeta}\brc{(\cD_{\ualp}^{[1]}\Phi)T^{[1]\ualp}
 +\frac{1}{4}\Phi\cD_{\ualp}^{[1]}T^{[1]\ualp}}, 
 \label{def:cT2}
\ee
where $T^{[1]\ualp}\equiv -\zeta T^{1\ualp}+T^{2\ualp}$. 
This certainly satisfies the condition~$\cD_{\ualp}^{[1]}\cT=0$ 
due to the constraints~(\ref{cstrt:tsrI}) and (\ref{cstrt:tsrII}). 
For an SU(2)-Majorana-Weyl spinor~$\Psi^{i\ualp}$, 
a quantity~$\Psi^{[1]\ualp}\equiv -\zeta\Psi^{1\ualp}+\Psi^{2\ualp}$ 
is transformed by the smile conjugation as
\be
 \Psi^{[1]\ualp} \to \breve{\Psi}^{[1]\ualp} \equiv 
 \left.\brkt{\overline{\Psi^{[1]}}}^{\ualp}\right|_{\zeta^*\to -1/\zeta} 
 = -\frac{1}{\zeta}\Psi^{[1]\ualp}, 
 \label{MWcond}
\ee
where the overline denotes the covariant conjugation defined by (\ref{def:ovln}), 
and we have used (\ref{SU2MWcond}). 
Using this property, it is shown 
that $\cT^{[2]}$ is smile-real ($\breve{\cT}^{[2]}=\cT^{[2]}$), 
and thus it can be the Lagrangian superfield for the tensor multiplets. 
\be
 L_{\rm tensor} = \cT^{[2]}. \label{L_tsr}
\ee

Besides the description by the tropical superfield, 
a 6D vector multiplet is also described by 
a superfield~$F^{i\ualp}$ subject to the same constraint as (\ref{cstrt:tsrII}) 
if it is further constrained by 
\bea
% \cD_{\ualp}^{(i}F^{j)\ubt}-\frac{1}{4}\dlt_{\ualp}^{\;\;\ubt}
% \cD_{\ugm}^{(i}F^{j)\ugm} \eql 0, \nonumber\\
%
 \cD_{i\ualp}F^{i\ualp} \eql 0.  \label{cstrt:F}
\eea
As we will see later, this superfield contains 
the field strength of the gauge field, and thus gauge-invariant. 
Using $F^{i\ualp}$ with $\Phi$ and $\bdm{V}$, 
we can construct the Lagrangian superfield for the vector-tensor couplings as  
\be
 L_{\rm VT} = \bdm{V}\cF^{[2]},  \label{L_VT}
\ee
where
\bea
 \cF^{[2]} \defa \frac{i}{\zeta}\brc{(\cD_{\ualp}^{[1]}\Phi)F^{[1]\ualp}
 +\frac{1}{4}\Phi\cD_{\ualp}^{[1]}F^{[1]\ualp}}, \nonumber\\
 F^{[1]\ualp} \defa -\zeta F^{1\ualp}+F^{2\ualp}.  
 \label{def:cF2}
\eea

The action constructed from the above Lagrangian superfields~(\ref{L_hyp}), (\ref{L_tsr}) 
and (\ref{L_VT}) is invariant under the following gauge transformations.  
\bea
 \dlt_\Lmd\bdm{V} \eql \bdm{\Lmd}+\breve{\bdm{\Lmd}}, \;\;\;\;\;
 \dlt_\Lmd\Ups = \bdm{\Lmd}\Ups, \;\;\;\;\;
 \dlt_\Lmd\Phi = \dlt_\Lmd T^{i\ualp} = 0, \nonumber\\
 \dlt_G T^{i\ualp} \eql \bdm{G}^{i\ualp}, \;\;\;\;\;
 \dlt_G \bdm{V} = \dlt_G\Ups = \dlt_G\Phi = 0, \label{gauge_trf}
\eea
where the transformation parameters~$\bdm{\Lmd}$ and $\bdm{G}^{i\ualp}$ 
are an arctic superfield and a constrained superfield that satisfies 
the same constraints as (\ref{cstrt:tsrII}) and (\ref{cstrt:F}), 
respectively.

\section{$N=1$ superfield description} \label{N1sf_action}
In this section, we express the $N=2$ invariant action 
in the previous section in terms of $N=1$ superfields. 
For this purpose, it is convenient to devide the bosonic coordinates~$x^M$ 
into the 4D part~$x^\mu$ ($\mu=0,1,2,3$) 
and the extra-dimensional part~$z\equiv\frac{1}{2}(x^4+ix^5)$ and $\bar{z}$. 
As for the fermionic coordinates~$\Tht^{i\ualp}$, they are decomposed into 
$(\tht^\alp,\bar{\tht}_{\dalp})$ that describes the $N=1$ subsuperspace we focus on 
and the rest part~$(\tht^{\prime\alp},\bar{\tht}'_{\dalp})$ 
as shown in (\ref{decomp:Tht}). 
We follow the notations of Ref.~\cite{Wess:1992cp} 
for the 2-component spinor indices.

\subsection{Superfield action formula}
Since the action~(\ref{action_fml}) is independent of the choice of $\eta$, 
we choose it as $\eta=0$ in the following. 
Then, $\cD^{[-4]}$ becomes 
\be
 \cD^{[-4]} = -\frac{1}{96}\ep^{\ualp\ubt\ugm\udlt}
 \cD_{\ualp}^1\cD_{\ubt}^1\cD_{\ugm}^1\cD_{\udlt}^1 
 = \frac{1}{16}D^2\bar{D}^{\prime 2}, 
 \label{expr:cD4}
\ee
where $D_\alp$ and $\bar{D}^{\prime\dalp}$ are defined in (\ref{def:DbD}), 
and we have used (\ref{decomp:ep}) and (\ref{decomp:cD}). 
Since the Lagrangian superfield~$L$ is a projective superfield, 
it satisfies $\cD_{\ualp}^{[1]}L=0$. 
From (\ref{decomp:cD}), this is rewritten as
\be
 \brkt{-\zeta D_\alp-D'_\alp}L = 0, \;\;\;\;\;
 \brkt{-\zeta\bar{D}^{\prime\dalp}+\bar{D}^{\dalp}}L = 0. 
\ee
Thus, $\cD^{[-4]}L$ is rewritten as 
\be
 \cD^{[-4]}L =\frac{1}{16}D^2\brkt{\frac{1}{\zeta}\bar{D}'\bar{D}L}
 = \frac{1}{16}D^2\brkt{\frac{1}{\zeta^2}\bar{D}^2-\frac{4}{\zeta}\der}L,  
\ee
where $\der\equiv\der_z=\der_4-i\der_5$, and we have used (\ref{D:formulae}). 
Therefore, the action~(\ref{action_fml}) becomes 
\bea
 S \eql \int\dr^6x\brc{\left.\oint_C\frac{d\zeta}{2\pi i\zeta}\frac{1}{16}D^2\bar{D}^2L
 \right|_{\tht=\tht'=0}} \nonumber\\
 \eql \int\dr^6x\brc{\oint_C\frac{d\zeta}{2\pi i\zeta}
 \int\dr^4\tht\;L|} 
 \equiv \int\dr^6x\;\cL.  \label{action_fml2}
\eea
where a total derivative term is dropped, 
and the symbol~$|$ denotes the projection~$\tht'=0$. 

For a given projective superfield~$\Xi(x,\Tht,\zeta)$, 
its expansion coefficients~$\Xi_n(x,\Tht)$ in (\ref{expand:Xi}) satisfy 
\be
 D_\alp\Xi_n = -D'_\alp\Xi_{n+1}, \;\;\;\;\;
 \bar{D}_{\dalp}\Xi_n = \bar{D}'_{\dalp}\Xi_{n-1}, \label{cstrt:Xi2}
\ee
which comes from the constraint~(\ref{cstrt:Xi}). 
Note that $\Xi_n(x,\Tht)$ is decomposed into 
the following $N=1$ superfields. 
\bea
 &&\Xi_n|, \;\;\;\;\;
 D'_\alp\Xi_n|, \;\;\;\;\;
 \bar{D}'_{\dalp}\Xi_n|, \;\;\;\;\;
 D^{\prime 2}\Xi_n|, \;\;\;\;\;
 D'_\alp\bar{D}'_{\dalp}\Xi_n|, \nonumber\\
 &&\bar{D}^{\prime 2}\Xi_n|, \;\;\;\;\;
 \bar{D}^{\prime 2}D'_\alp\Xi_n|, \;\;\;\;\;
 D^{\prime 2}\bar{D}'_{\dalp}\Xi_n|, \;\;\;\;\;
 D^{\prime 2}\bar{D}^{\prime 2}\Xi_n|. 
\eea
The condition~(\ref{cstrt:Xi2}) provides 
constraints on these $N=1$ superfields. 
The action formula~(\ref{action_fml2}) is expressed 
in terms of them. 
Although each projective superfield contains infinite number of $N=1$ superfields, 
only a finite small number of them survive in the final expression 
of the action as we will see below.

As a simple example, let us consider a free hypermultiplet. 
The Lagrangian is given by 
\be
 \cL_{\rm hyp} = \int\dr^4\tht\;\left.\oint_C\frac{d\zeta}{2\pi i\zeta}\;
 \breve{\Ups}\Ups\right| 
 = \int\dr^4\tht\;\left.\brc{\sum_{n=0}^\infty (-1)^n\abs{\Ups_n}^2}\right|.  
 \label{cL_hyp1}
\ee
Since the arctic superfield~$\Ups$ does not have terms with negative power in $\zeta$ 
(\ie, $\Ups_n=0$ for $n<0$), the constraint~(\ref{cstrt:Xi2}) becomes 
\bea
 D_\alp\Ups_n \eql -D'_\alp\Ups_{n+1} \;\;\; (n\geq 0), \;\;\;\;\;
 D'_\alp\Ups_0 = 0, \nonumber\\
 \bar{D}_{\dalp}\Ups_n \eql \bar{D}'_{\dalp}\Ups_{n-1} 
 \;\;\; (n\geq 1), \;\;\;\;\;
 \bar{D}_{\dalp}\Ups_0 = 0.  \label{cstrt:Ups}
\eea
Thus the constraints on $\Ups_0|$ and $\Ups_1|$ are isolated 
from the other $N=1$ superfields. 
\be
 \bar{D}_{\dalp}\Ups_0| = 0, \;\;\;\;\;
 \bar{D}^2\Ups_1| = 4\der\Ups_0|. 
\ee
We have used (\ref{D:formulae}). 
Note that $\Ups_n|$ ($n\geq 2$) are unconstrained superfields.\footnote{
From (\ref{cstrt:Ups}), each $\Ups_n|$ ($n\geq 2$) is related to $D'\Ups_{n\pm 1}|$. 
However, since the latter does not appear in the action, 
the former can be regarded as an unconstrained superfield. 
}  
Hence they can be easily integrated out and obtain 
\be
 \cL_{\rm hyp} = \int\dr^4\tht\;\left.\brc{\abs{\Ups_0}^2-\abs{\Ups_1}^2}\right|. 
 \label{cL_hyp2}
\ee
This is further rewritten as 
\be
 \cL_{\rm hyp} = \int\dr^4\tht\;\sbk{\brkt{\abs{\Phi}^2-\abs{\xi}^2}
 +\brc{\kp\brkt{\bar{D}^2\xi-4\der\Phi}+\hc}}, 
\ee
where $\Phi\equiv\Ups_0|$ is a chiral superfield, 
and $\xi$ and $\kp$ are unconstrained $N=1$ superfields. 
In fact, integrating out $\kp$ and $\bar{\kp}$, 
this reduces to (\ref{cL_hyp2}) with $\xi=\Ups_1|$. 
On the other hand, if we integrate out $\xi$ and $\bar{\xi}$, 
we obtain~\cite{Gates:2005mc}
\bea
 \cL_{\rm hyp} \eql \int\dr^4\tht\;\brc{\abs{\Phi}^2
 +|\bar{D}^2\kp|^2-\brkt{4\kp\der\Phi+\hc}} \nonumber\\
 \eql \int\dr^4\tht\brkt{\abs{\Phi}^2+|\tl{\Phi}|^2}
 +\brc{\int\dr^2\tht\;\tl{\Phi}\der\Phi+\hc}, 
\eea
where $\tl{\Phi}\equiv\bar{D}^2\kp$ is another chiral superfield, 
up to total derivatives. 
This is consistent with (2.3) in Ref.~\cite{ArkaniHamed:2001tb}. 

\ignore{
Each component superfield~$\Ups_n|$ is transformed under 
the gauge transformation~(\ref{gauge_trf}) as
\be
 \Ups_n| \to \Ups'_n| = \sum_{k=0}^n a_k\Ups_{n-k}|, 
\ee
where $a_k$ are $N=1$ superfields 
defined as $e^{\sbdm{\Lmd}}|=\sum_{k=0}^\infty a_k\zeta^k$. 
Notice that $a_k$ ($k\geq 2$) are unconstrained $N=1$ superfields. 
Hence we can choose them so that
\be
 \Ups'_n| = 0. \;\;\;\;\; (n\geq 2)
\ee
In this gauge, the Lagrangian in this sector~(\ref{L_hyp}) becomes 
\bea
 \cL_{\rm hyp} \eql \int\dr^4\tht\oint_C\frac{d\zeta}{2\pi i\zeta}
 \left.\breve{\Ups}e^{-\sbdm{V}}\Ups\right| \nonumber\\
 \eql \int\dr^4\tht\;\left.\brc{b_0\brkt{\abs{\Ups_0}^2-\abs{\Ups_1}^2}
 +\brkt{-b_1\bar{\Ups}_1\Ups_0+\hc}}\right| 
 \label{cL_hyp}
\eea
where
\be
 e^{-\sbdm{V}}| = \sum_{k=-\infty}^\infty b_k\zeta^k, \;\;\;\;\;
 b_{-k} = (-1)^k\bar{b}_k. 
\ee
Since $\Ups_0|$ and $\Ups_1|$ are constrained $N=1$ superfields, 
(\ref{cL_hyp}) is rewritten as~\cite{Gates:2005mc}
\bea
 \cL_{\rm hyp} \eql \int\dr^4\tht\;\left[
 b_0\brkt{\abs{\Phi}^2-\abs{\xi}^2}-\brkt{b_1\bar{\xi}\Phi+\hc} \right.\nonumber\\
 &&\hspace{10mm}\left.
 +\brc{\eta\brkt{\bar{D}^2\xi-4\der\Phi}+\hc}\right], 
\eea
where $\Phi\equiv\Ups_0|$ is a chiral superfield, 
and $\xi$ and $\eta$ are unconstrained superfield. 
In fact, integrating out $\eta$ and $\bar{\eta}$, this reduces to (\ref{cL_hyp}) 
with $\xi=\Ups_1|$. 
On the other hand, if we integrate out $\xi$ and $\bar{\xi}$, we obtain
\bea
 \cL_{\rm hyp} \eql \int\dr^4\tht\;\brc{
 b_0\abs{\Phi}^2+\frac{1}{b_0}\abs{D^2\bar{\eta}-b_1\Phi}^2
 -\brkt{4\eta\der\Phi+\hc}} \nonumber\\
 \eql 
\eea
}

\subsection{Decomposition into $N=1$ superfields}
The constraint~(\ref{cstrt:tsrII}) is rewritten as
\bea
% \cD_{\ualp}^{[1]}\cD_{\ubt}^{[1]}\Phi \eql 0, \nonumber\\
%
 \cD_{\ualp}^{[1]}T^{[1]\ubt}-\frac{1}{4}\dlt_{\ualp}^{\;\;\ubt}
 \cD_{\ugm}^{[1]}T^{[1]\ugm} \eql 0. 
\eea
Since $\brc{\cD_{\ualp}^{[1]},\cD_{\ubt}^{[1]}}=0$, 
the solution of this constraint is expressed as~\cite{Sokatchev:1988aa}
\bea
% \Phi \eql \frac{1}{3!\zeta^2}\ep^{\ualp\ubt\ugm\udlt}
% \cD_{\ualp}^{[1]}\cD_{\ubt}^{[1]}\cD_{\ugm}^{[1]}
% \brkt{-\zeta P^1_{\udlt}+P^2_{\udlt}}, \nonumber\\
%
 T^{[1]\ualp} \eql \frac{i}{3!\zeta}\ep^{\ualp\ubt\ugm\udlt}
 \cD_{\ubt}^{[1]}\cD_{\ugm}^{[1]}\cD_{\udlt}^{[1]}\bdm{P}^{[-2]}, 
 \label{sol:T1}
\eea
where the prepotential~$\bdm{P}^{[-2]}$ is a $\zeta$-independent $N=2$ superfield, 
which is a real scalar. 
The overall $\zeta$-dependence is determined so that $T^{[1]\ualp}$ satisfy
\be
% \breve{\Phi} = \Phi, \;\;\;\;\;
%
 \breve{T}^{[1]\ualp} = -\frac{1}{\zeta}T^{[1]\ualp}. 
\ee
(See (\ref{MWcond}).)
In the 2-component-spinor notation, (\ref{sol:T1}) is rewritten as 
\bea
 T^{[1]\alp} \eql \frac{i}{2\zeta}\brkt{-\zeta\bar{D}'+\bar{D}}^2
 \brkt{-\zeta D^\alp-D^{\prime\alp}}\bdm{P}^{[-2]}, \nonumber\\
 T^{[1]}_{\dalp} \eql \frac{i}{2\zeta}\brkt{-\zeta D-D'}^2
 \brkt{-\zeta\bar{D}'_{\dalp}+\bar{D}_{\dalp}}\bdm{P}^{[-2]}. 
\eea
We have used (\ref{decomp:ep}) and (\ref{decomp:cD}). 

Since $T^{[1]\ualp}$ is a linear function of $\zeta$, 
the prepotential~$\bdm{P}^{[-2]}$ should satisfy
\be
 \bar{D}_{\dalp}D^{\prime 2}\bdm{P}^{[-2]} = \bar{D}^2D'_\alp\bdm{P}^{[-2]} = 0. 
 \label{cstrt:P_T}
\ee
From the linear and constant terms in $\zeta$, we can read off 
the components~$T^{i\ualp}$ as 
\bea
 T^{1\alp} \eql \frac{i}{2}\brkt{D^{\prime\alp}\bar{D}^{\prime 2}
 -2D^\alp\bar{D}\bar{D}'+4\der D^\alp}\bdm{P}^{[-2]}, 
 \nonumber\\
 T^{2\alp} \eql -\frac{i}{2}\brkt{\bar{D}^2D^\alp-2\bar{D}\bar{D}'D^{\prime\alp}
 +4\der D^{\prime\alp}}\bdm{P}^{[-2]},  \nonumber\\
 T^1_{\dalp} \eql -(T^2_\alp)^*, \;\;\;\;\;
 T^2_{\dalp} = (T^1_\alp)^*.  \label{comp:T^i}
\eea
Then, $\Phi$ constructed by (\ref{rel:Phi-T}) is calculated as
\bea
 \Phi \eql D_\alp T^{2\alp}+\bar{D}^{\prime\dalp}T^2_{\dalp}
 +D'_\alp T^{1\alp}-\bar{D}^{\dalp}T^1_{\dalp} \nonumber\\
 \eql D_\alp T^{2\alp}+D'_\alp T^{1\alp}+\hc \nonumber\\
 \eql i\brkt{-2D^\alp\bar{D}\bar{D}'D'_\alp
 +2\bar{D}_{\dalp}DD'\bar{D}^{\prime\dalp}
 +4\der DD'-4\bar{\der}\bar{D}\bar{D}'}\bdm{P}^{[-2]}.  \label{expr:Phi}
\eea

The $N=2$ superfield~$\bdm{P}^{[-2]}$ is decomposed into 
the following $N=1$ superfields. 
\bea
 p_0 \defa \bdm{P}^{[-2]}|, \;\;\;\;\;
 p_1^\alp \equiv D^{\prime\alp}\bdm{P}^{[-2]}|, \nonumber\\
 p_2^{\dalp\alp} \defa \bar{D}^{\prime\dalp}D^{\prime\alp}\bdm{P}^{[-2]}|, \;\;\;\;\;
 p_3 \equiv D^{\prime 2}\bdm{P}^{[-2]}|, \nonumber\\
 p_4^\alp \defa D^{\prime\alp}\bar{D}^{\prime 2}\bdm{P}^{[-2]}|, \;\;\;\;\;
 p_5 \equiv D^{\prime\alp}\bar{D}^{\prime 2}D'_\alp \bdm{P}^{[-2]}|. 
 \label{def:ps}
\eea
Then, (\ref{cstrt:P_T}) is translated 
into the following constraints.\footnote{
The last four constraints are obtained by operating $D'_\bt$ or$\bar{D}'_{\dbt}$ 
on (\ref{cstrt:P_T}) and putting $\tht'=0$. }
\bea
 &&\bar{D}^2p_1^\alp = 0, \;\;\;\;\;
 \bar{D}_{\dalp}p_3 = 0, \nonumber\\
 &&\bar{D}_{\dalp}\bar{p}_{4\dbt}+2\ep_{\dalp\dbt}\der p_3 = 0, \;\;\;\;\;
 \bar{D}^2p_2^{\dalp\alp}+4\der\bar{D}^{\dalp}p_1^\alp = 0, \nonumber\\
 &&\bar{D}_{\dalp}p_5-4i\sgm^\mu_{\alp\dbt}\der_\mu\bar{D}_{\dalp}p_2^{\dbt\alp}
 +4\der\bar{p}_{4\dalp} = 0, \nonumber\\
 &&\bar{D}^2\brkt{p_4^\alp-4i\bar{\sgm}^{\mu\dalp\alp}\der_\mu\bar{p}_{1\dalp}}
 +8\der\bar{D}_{\dalp}p_2^{\dalp\alp}-16\der^2 p_1^\alp = 0. 
 \label{cstrt:ps}
\eea
When the spinor derivatives~$D_\alp$ and $\bar{D}^{\dalp}$ 
act on $N=1$ superfields, they are understood as the 4D $N=1$ ones, 
\ie, $D_\alp=\frac{\der}{\der\tht^\alp}+i(\sgm^\mu\bar{\tht})_\alp\der_\mu$ 
and $\bar{D}^{\dalp}=\frac{\der}{\der\bar{\tht}_{\dalp}}
+i(\bar{\sgm}^\mu\tht)^{\dalp}\der_\mu$.  
From the second and the third constraints in (\ref{cstrt:ps}), we obtain
\be
 \bar{D}^2\bar{p}_4^{\dalp} = 0. 
\ee
Namely, $p_4^\alp$ is a complex anti-linear superfield and expressed as
\be
 p_4^\alp = D^\alp q_4, 
\ee
where $q_4$ is a complex scalar superfield. 
The fourth constraint in (\ref{cstrt:ps}) indicates that
\be
 \chi^\alp \equiv \bar{D}_{\dalp}p_2^{\dalp\alp}-2\der p_1^\alp 
\ee
is a chiral superfield. 
Thus $\chi^\alp$ can be expressed as $\chi^\alp=\bar{D}^2U_\chi^\alp$, 
where $U_\chi^\alp$ is a spinor superfield. 
The sixth constraint in (\ref{cstrt:ps}) is rewritten as
\bea
 0 \eql \bar{D}^2\brkt{p_4^\alp+2\brc{D^\alp,\bar{D}^{\dalp}}\bar{p}_{1\dalp}}
 +8\der\chi^\alp \nonumber\\
 \eql \bar{D}^2\brkt{p_4^\alp+2D^\alp\bar{D}^{\dalp}\bar{p}_{1\dalp}+8\der U_\chi^\alp}, 
\eea
which indicates that
\be
 Z^\alp \equiv \frac{1}{2}p_4^\alp-D^\alp\bar{D}_{\dalp}\bar{p}_1^{\dalp}+4\der U_\chi^\alp
 \label{def:Z}
\ee
is a complex linear superfield, \ie, $\bar{D}^2Z^\alp=0$.

\subsection{$N=1$ description of tensor multiplet} \label{N1tsr}
From (\ref{comp:T^i}) and (\ref{def:ps}), we obtain
\bea
 T^{1\alp}| \eql \frac{i}{2}\brkt{
 p_4^\alp-2D^\alp\bar{D}_{\dalp}\bar{p}_1^{\dalp}
 +4\der D^\alp p_0} 
 = iD^\alp X, \nonumber\\
 T^{2\alp}| \eql -\frac{i}{2}\brkt{\bar{D}^2D^\alp p_0
 -2\bar{D}_{\dalp}\bar{p}_2^{\dalp\alp}+4\der p_1^\alp} 
 = i\bar{D}^2Y^\alp, 
 \label{expr:T^i}
\eea
where
\bea
 X \defa \frac{1}{2}q_4-\bar{D}_{\dalp}\bar{p}_1^{\dalp}+2\der p_0, \nonumber\\
 Y^\alp \defa U_\chi^\alp-\frac{1}{2}D^\alp p_0. 
\eea
Using $Z^\alp$ defined in (\ref{def:Z}), (\ref{expr:T^i}) is also expressed as
\be
 T^{1\alp}| = i\brkt{Z^\alp-4\der Y^\alp}, \;\;\;\;\;
 T^{2\alp}| = i\bar{D}^2Y^\alp.  \label{expr:T^i2}
\ee
From (\ref{comp:T^i}), we can calculate
\bea
 \cD^{[1]}_{\ualp}T^{[1]\ualp} \eql 
 \brkt{-\zeta D_\alp-D'_\alp}T^{[1]\alp}
 +\brkt{-\zeta\bar{D}^{\prime\dalp}+\bar{D}^{\dalp}}T^{[1]}_{\dalp} \nonumber\\
 \eql i\left\{\zeta^2\brkt{-4\der D^2+\frac{3}{2}D^2\bar{D}\bar{D}'
 -2i\sgm^\mu_{\alp\dalp}\der_\mu D^\alp\bar{D}^{\prime\dalp}
 +2\bar{\der}\bar{D}^{\prime 2}-\frac{3}{2}DD'\bar{D}^{\prime 2}} \right.\nonumber\\
 &&\hspace{5mm}
 +\zeta\left(-D^\alp\bar{D}^2D_\alp+16\der\bar{\der}-8\der DD'
 -8\bar{\der}\bar{D}\bar{D}' \right.\nonumber\\
 &&\hspace{15mm}\left.
 +2D^\alp\bar{D}\bar{D}'D'_\alp+2\bar{D}_{\dalp}DD'\bar{D}^{\prime\dalp}
 -D^{\prime\alp}\bar{D}^{\prime 2}D'_\alp\right) \nonumber\\
 &&\hspace{5mm}\left.
 +\brkt{4\bar{\der}\bar{D}^2-\frac{3}{2}\bar{D}^2DD'
 +2i\sgm^\mu_{\alp\dalp}\der_\mu\bar{D}^{\dalp}D^{\prime\alp}
 -2\der D^{\prime 2}+\frac{3}{2}\bar{D}\bar{D}'D^{\prime 2}}\right\}
 \bdm{P}^{[-2]}. \nonumber\\
\eea
Thus, 
\be
 \left.\frac{i}{\zeta}\cD_{\ualp}^{[1]}T^{[1]\ualp}\right| 
 = \zeta\cA_1+\cA_0-\frac{1}{\zeta}\bar{\cA}_1, 
\ee
where 
\bea
 \cA_1 \eql 4\der D^2p_0-\frac{3}{2}D^2\bar{D}_{\dalp}\bar{p}_1^{\dalp}
 -\frac{1}{2}\sbk{D^2,\bar{D}_{\dalp}}\bar{p}_1^{\dalp}
 -2\bar{\der}\bar{p}_3+\frac{3}{2}D^\alp p_{4\alp} \nonumber\\
 \eql D^\alp\brkt{p_{4\alp}-2D_\alp\bar{D}_{\dalp}\bar{p}_1^{\dalp}
 +4\der D_\alp p_0} \nonumber\\
 \eql 2D^\alp\brkt{Z_\alp-4\der Y_\alp}, \nonumber\\
 \cA_0 \eql \brkt{D^\alp\bar{D}^2D_\alp-16\der\bar{\der}}p_0
 +8\der D^\alp p_{1\alp}+8\bar{\der}\bar{D}_{\dalp}\bar{p}_1^{\dalp} \nonumber\\
 &&-2D^\alp\bar{D}_{\dalp}(p_2)^{\dalp}_{\;\;\alp}
 -2\bar{D}_{\dalp}D^\alp(\bar{p}_2)_\alp^{\;\;\dalp}
 +p_5. 
\eea
Although $\cA_0$ cannot be expressed in terms of only $Y_\alp$ and $Z_\alp$, 
$\bar{D}_{\dalp}\cA_0$ can. 
In fact, after some calculations, we obtain 
\bea
 \bar{D}_{\dalp}\cA_0 \eql -8\der\brkt{\bar{Z}_{\dalp}-4\bar{\der}\bar{Y}_{\dalp}}
 +2\bar{D}^2D^2\bar{Y}_{\dalp} \nonumber\\
 \eql \bar{D}_{\dalp}\brkt{-8\der\bar{X}-4\bar{D}_{\dbt}D^2\bar{Y}^{\dbt}}, 
\eea
where we have used that $Z^\alp-4\der Y^\alp = D^\alp X (= -iT^{1\alp}|)$. 
Thus, $\cA_0$ is expressed as
\be
 \cA_0 = -8\der\bar{X}-4\bar{D}_{\dalp}D^2\bar{Y}^{\dalp}+\phi_0, 
\ee
where $\phi_0$ is a chiral superfield that is determined so that $\cA_0$ is real. 

From (\ref{expr:Phi}), we have
\bea
 \Phi| \eql i\brkt{-2D^\alp\bar{D}_{\dalp}(p_2)^{\dalp}_{\;\;\alp}
 +2\bar{D}_{\dalp}D^\alp(\bar{p}_2)_\alp^{\;\;\dalp}
 +4\der D^\alp p_{1\alp}-4\bar{\der}\bar{D}_{\dalp}\bar{p}_1^{\dalp}} \nonumber\\
 \eql -2iD^\alp\chi_\alp+2i\bar{D}_{\dalp}\bar{\chi}^{\dalp} \nonumber\\
% \eql -2iD^\alp\bar{D}^2\brkt{Y_\alp+\frac{1}{2}D_\alp p_0}
% +2i\bar{D}_{\dalp}D^2\brkt{\bar{Y}^{\dalp}+\frac{1}{2}\bar{D}^{\dalp}p_0} \nonumber\\
 \eql -2iD^\alp\bar{D}^2Y_\alp+2i\bar{D}_{\dalp}D^2\bar{Y}^{\dalp}, \nonumber\\
 D'_\alp\Phi| \eql -\frac{i}{2}\bar{D}^2D^2p_{1\alp}
 +\frac{i}{2}\brkt{D_\alp\bar{D}_{\dalp}
 +2\bar{D}_{\dalp}D_\alp}\bar{p}_4^{\dalp}-4i\bar{\der}\chi_\alp \nonumber\\
 \eql i\brkt{D_\alp\bar{D}_{\dalp}+2\bar{D}_{\dalp}D_\alp}
 \brkt{\bar{Z}^{\dalp}-4\bar{\der}\bar{Y}^{\dalp}}-4i\bar{\der}\bar{D}^2Y_\alp. 
\eea

In summary, the 6D tensor multiplet is described by 
the spinor superfields~$Y_\alp$ and $Z_\alp$, 
where the latter is constrained by $\bar{D}^2Z_\alp=0$.

\subsection{$N=1$ description of vector multiplet}
As mentioned in Sec.~\ref{VTsector}, a 6D vector multiplet can also be 
described by the constrained superfield~$F^{i\ualp}$. 
This is decomposed into $N=1$ superfields in a similar way to the tensor multiplet. 
\bea
 F^{1\alp}| \eql i\brkt{Z_F^\alp-4\der Y_F^\alp} 
 = iD^\alp X_F, \;\;\;\;\;
 F^{2\alp}| = i\bar{D}^2Y_F^\alp, \nonumber\\
 \left.\frac{i}{\zeta}\cD_{\ualp}^{[1]}F^{[1]\ualp}\right| \eql 
 \zeta\cA_{F1}+\cA_{F0}-\frac{1}{\zeta}\cA_{1F}^*, \nonumber\\
 \cA_{F1} \eql 2D^\alp\brkt{Z_{F\alp}-4\der Y_{F\alp}}, \nonumber\\
 \cA_{F0} \eql -8\der\bar{X}_F-4\bar{D}_{\dalp}D^2\bar{Y}_F^{\dalp}+\phi_{F0}, 
 \nonumber\\
 \cD_{i\ualp}F^{i\ualp}| \eql -2iD^\alp\bar{D}^2Y_{F\alp}
 +2i\bar{D}_{\dalp}D^2\bar{Y}_F^{\dalp}, 
 \label{expr:vectorSF}
\eea
where $Z_F^\alp$ and $\phi_{F0}$ 
are a complex linear and a chiral superfields, respectively. 
In contrast to the tensor multiplet, $F^{i\ualp}$ is further 
constrained by (\ref{cstrt:F}), which indicates that 
\be
 D^\alp\bar{D}^2Y_{F\alp} = \bar{D}_{\dalp}D^2\bar{Y}_F^{\dalp}. 
\ee
This is regarded as the Bianchi identity, and solved as 
\be
 Y_F^\alp = -\frac{1}{4}D^\alp V, 
\ee
where $V$ is an unconstrained real superfield. 
Then, $Z_F^\alp$ is expressed as
\be
 Z_F^\alp = 4\der Y_F^\alp+D^\alp X_F = -D^\alp\Sgm, 
\ee
where $\Sgm\equiv \der V-X_F$. 
Since $\bar{D}^2Z_F^\alp=0$, $\Sgm$ is a chiral superfield.  
Thus, (\ref{expr:vectorSF}) becomes
\bea
 F^{1\alp}| \eql iD^\alp\brkt{\der V-\Sgm}, \;\;\;\;\;
 F^{2\alp} = -\frac{i}{4}\bar{D}^2D^\alp V, \nonumber\\
 \cA_{F1} \eql 2D^2\brkt{\der V-\Sgm}, \;\;\;\;\;
 \cA_{F0} = -8\der\brkt{\bar{\der}V-\bar{\Sgm}}
 +\bar{D}_{\dalp}D^2\bar{D}^{\dalp}V+\phi_{F0}.  
 \label{expr:F^i}
\eea
As mentioned in Sec.~\ref{VTsector}, $F^{i\alp}$ are invariant 
under the gauge transformation, 
\be
 V \to V+\Lmd+\bar{\Lmd}, \;\;\;\;\;
 \Sgm \to \Sgm+\der \Lmd, 
\ee
where $\Lmd$ is a chiral superfield. 
Especially, $F^{2\alp}$ is proportional to 
the field strength superfield, 
\be
 \cW^\alp \equiv -\frac{1}{4}\bar{D}^2D^\alp V. 
\ee
Since $\phi_{F0}$ is a chiral superfield and $\cA_{F0}$ is real, 
we find that $\phi_{F0}=8\bar{\der}\Sgm$. 
Namely, 
\be
 \cA_{F0} = 8\brc{-\brkt{\Box_4\cPT+\der\bar{\der}}V+\bar{\der}\Sgm+\der\bar{\Sgm}}, 
\ee
where $\Box_4\equiv\der_\mu\der^\mu$, and 
\be
 \cPT \equiv -\frac{\bar{D}_{\dalp}D^2\bar{D}^{\dalp}}{8\Box_4}  
\ee
is the projection operator~\cite{Wess:1992cp}. 

In summary, the 6D vector multiplet is described by 
a chiral superfield~$\Sgm$ and a real superfield~$V$, 
which are independent of each other.

\subsection{Vector-tensor couplings}
Now we consider the vector-tensor couplings. 
Note that $\cF^{[2]}$ defined in (\ref{def:cF2}) is an O(2) multiplet, \ie, 
\be
 \cF^{[2]} = \zeta\cF_1^{[2]}+\cF_0^{[2]}-\frac{1}{\zeta}(\cF_1^{[2]})^*, 
\ee
where $\cF_0^{[2]}$ is real. 
Since 
\bea
 \frac{i}{\zeta}\brkt{\cD_{\ualp}^{[1]}\Phi}F^{[1]\ualp} 
 \eql i\zeta\brc{D_\alp\Phi F^{1\alp}-(D'_\alp\Phi F^{2\alp})^*} \nonumber\\
 &&+i\brc{D'_\alp\Phi F^{1\alp}-D_\alp\Phi F^{2\alp}
 -(D'_\alp\Phi F^{1\alp})^*+(D_\alp\Phi F^{2\alp})^*} \nonumber\\
 &&-\frac{i}{\zeta}\brc{D'_\alp\Phi F^{2\alp}-(D_\alp\Phi F^{1\alp})^*}, 
\eea
we can calculate $\cF_1^{[2]}|$ and $\cF_0^{[2]}|$ 
after some calculations 
by using the results in the previous subsections as 
\bea
 \cF_1^{[2]}| \eql \frac{1}{2}D^2\brc{\Phi_T\brkt{\der V-\Sgm}}
 -\bar{\cW}_T\bar{\cW}, \nonumber\\
 \cF_0^{[2]}| \eql \brc{-\cW_T^\alp
 D_\alp\brkt{\der V-\Sgm}-D^\alp\Phi_T\cW_\alp+\hc} \nonumber\\
 &&-2\Phi_T\brc{\brkt{\Box_4\cPT+\der\bar{\der}}V
 -\bar{\der}\Sgm-\der\bar{\Sgm}}, 
 \label{expr:cF2i}
\eea
where 
\bea
 \Phi_T \defa \Phi| = -2iD^\alp\bar{D}^2Y_\alp
 +2i\bar{D}_{\dalp}D^2\bar{Y}^{\dalp}, \nonumber\\
 \cW_{T\alp} \defa i\bar{D}^2\brkt{D_\alp\bar{X}+4\bar{\der}Y_\alp} \nonumber\\
 \eql -i\brkt{D_\alp\bar{D}_{\dalp}+2\bar{D}_{\dalp}D_\alp}
 \brkt{\bar{Z}^{\dalp}-4\bar{\der}\bar{Y}^{\dalp}}
 +4i\bar{\der}\bar{D}^2Y_\alp. 
 \label{def:PhiT-cWT}
\eea
Note that the real linear superfield~$\Phi_T$ and 
and the chiral superfield~$\cW_T^\alp$ are not independent.  
As shown in Appendix~\ref{cstrt:tsr}, they are related through 
\bea
 D^\alp\cW_{T\alp} \eql -2\bar{\der}\Phi_T, \nonumber\\
 \bar{D}^2D^\alp\Phi_T \eql -4\der\cW_T^\alp. \label{cstrt:Phi_T}
\eea
From these relations, we obtain
\be
 \brkt{\Box_4+\der\bar{\der}}\Phi_T = 0, \;\;\;\;\;
 \brkt{\Box_4+\der\bar{\der}}\cW_T^\alp = 0, 
\ee
where we have used that $\cPT\Phi_T=\Phi_T$ 
and $\bar{D}^2D^2\cW_T^\alp=16\Box_4\cW_T^\alp$. 
Namely, $\Phi_T$ and $\cW_T^\alp$ are on-shell. 
This stems from the fact that the 6D tensor multiplet contains 
a self-dual tensor field~$B_{\mu\nu}^+$. 
In the 6D global SUSY theories, the tensor multiplet 
cannot be described as off-shell superfields,\footnote{
This fact is explicitly shown in the harmonic superspace formulation 
in Ref.~\cite{Sokatchev:1988aa}.} 
and thus should be treated as external fields. 
As shown in Ref.~\cite{Bergshoeff:1985mz,Coomans:2011ih}, 
the off-shell description of the tensor multiplet 
becomes possible by combining the Weyl multiplet 
when the theory is promoted to SUGRA. 

Therefore, from (\ref{L_VT}), the Lagrangian in the vector-tensor sector is 
\bea
 \cL_{\rm VT} \eql \oint_C\frac{d\zeta}{2\pi i\zeta}\int\dr^4\tht\;L_{\rm VT}| 
 \nonumber\\
 \eql \int\dr^4\tht\left.
 \brc{-V_1(\cF_1^{[2]})^*+V_0\cF_0^{[2]}-\bar{V}_1\cF_1^{[2]}}\right| \nonumber\\
 \eql \int\dr^4\tht\left[
 -\frac{1}{2}V_1|\bar{D}^2\brc{\Phi_T\brkt{\bar{\der} V-\bar{\Sgm}}}
 +V_1|\cW_T\cW \right.\nonumber\\
 &&\hspace{12mm}
 -V_0|\brc{\cW_T^\alp D_\alp\brkt{\der V-\Sgm}
 +D^\alp\Phi_T\cW_\alp+\hc} \nonumber\\
 &&\hspace{12mm}
 -2V_0|\Phi_T\brc{\brkt{\Box_4\cPT+\der\bar{\der}}V-\bar{\der}\Sgm-\der\bar{\Sgm}}
 \nonumber\\
 &&\hspace{12mm}\left.
 -\frac{1}{2}\bar{V}_1|D^2\brc{\Phi_T\brkt{\der V-\Sgm}}
 +\bar{V}_1|\bar{\cW}_T\bar{\cW}\right],
\eea
where $V_0$ and $V_1$ are the coefficient superfields 
in the tropical superfield~(\ref{tropical}). 
Using $d^2\bar{\tht}=-\frac{1}{4}\bar{D}^2$ and performing the partial integrals, 
the above Lagrangian is rewritten as
\bea
 \cL_{\rm VT} \eql -\int\dr^2\tht\;\tl{\Sgm}\cW_T\cW+\hc \nonumber\\
 &&-\int\dr^4\tht\left\{2\bar{\tl{\Sgm}}\Phi_T\brkt{\der V-\Sgm}
 +\tl{V}\cW_T^\alp D_\alp\brkt{\der V-\Sgm}
 +\tl{V}D^\alp\Phi_T\cW_\alp+\hc\right\} \nonumber\\
 &&-\int\dr^4\tht\;2\Phi_T\tl{V}\brc{\brkt{\Box_4\cPT+\der\bar{\der}}V
 -\bar{\der}\Sgm-\der\bar{\Sgm}} \label{cL_VT}
\eea
where
\be
 \tl{V} \equiv V_0|, \;\;\;\;\;
 \tl{\Sgm} \equiv \frac{1}{4}\bar{D}^2V_1|. 
\ee
The second line in (\ref{cL_VT}) is further rewritten as
\bea
 &&-\int\dr^4\tht\left\{2\bar{\tl{\Sgm}}\Phi_T\brkt{\der V-\Sgm}
 +\tl{V}\cW_T^\alp D_\alp\brkt{\der V-\Sgm}
 +\tl{V}D^\alp\Phi_T\cW_\alp+\hc\right\} \nonumber\\
 \eql -\int\dr^2\tht\brc{\Sgm\tl{\cW}\cW_T
 +\frac{1}{4}\bar{D}^2\brkt{\Phi_TD^\alp\tl{V}\cW_\alp
 +\der VD^\alp \tl{V}\cW_{T\alp}}}+\hc \nonumber\\
 &&+\int\dr^4\tht\;\Phi_T\tl{V}\brc{4\brkt{\Box_4\cPT+\der\bar{\der}}V
 -2\bar{\der}\Sgm-2\der\bar{\Sgm}} \nonumber\\
  &&+\int\dr^4\tht\brc{2\Phi_T(\bar{\der}\tl{V}-\bar{\tl{\Sgm}})
 \brkt{\der V-\Sgm}+\hc}, 
\eea
where we have used (\ref{cstrt:Phi_T}). 
Thus, $\cL_{\rm VT}$ becomes
\bea
 \cL_{\rm VT} \eql -\int\dr^2\tht\;\brc{\brkt{\tl{\Sgm}\cW+\Sgm\tl{\cW}}\cW_T
 +\frac{1}{4}\bar{D}^2\brkt{\Phi_TD^\alp\tl{V}\cW_\alp
 +\der VD^\alp\tl{V}\cW_{T\alp}}}+\hc \nonumber\\
 &&+\int\dr^4\tht\;2\Phi_T\tl{V}\brkt{\Box_4\cPT+\der\bar{\der}}V \nonumber\\
 &&+\int\dr^4\tht\brc{2\Phi_T(\bar{\der}\tl{V}-\bar{\tl{\Sgm}})
 \brkt{\der V-\Sgm}+\hc}, \label{expr:cL_VT}
\eea
where $\tl{\cW}_\alp\equiv -\frac{1}{4}\bar{D}^2D_\alp\tl{V}$. 
When the 6D vector multiplets~$(V,\Sgm)$ and $(\tl{V},\tl{\Sgm})$ are identical, 
(\ref{expr:cL_VT}) is simplified as
\bea
 \cL_{\rm VT} \eql -\int\dr^2\tht\;\brc{2\Sgm\cW\cW_T
 +\frac{1}{4}\bar{D}^2\brkt{\Phi_TD^\alp V\cW_\alp+\der VD^\alp V\cW_{T\alp}}}+\hc 
 \nonumber\\
 &&+\int\dr^4\tht\;2\Phi_T\brc{V\brkt{\Box_4\cPT+\der\bar{\der}}V
 +2\brkt{\bar{\der}V-\bar{\Sgm}}\brkt{\der V-\Sgm}}. 
 \label{expr:cL_VT2}
\eea
This is our main result. 
This contains the result in Ref.~\cite{ArkaniHamed:2001tb}  
as a special case:~$\Phi_T=1$ and $\cW_T^\alp=0$, 
which corresponds to the case where the tensor multiplet is absent. 
In such a case, (\ref{expr:cL_VT2}) becomes
\bea
 \cL_{\rm VT} \eql \int\dr^2\tht\;\cW^2+\hc \nonumber\\
 &&+\int\dr^4\tht\;\brc{VD^\alp\cW_\alp+2V\der\bar{\der}V
 +4\brkt{\bar{\der}V-\bar{\Sgm}}\brkt{\der V-\Sgm}} \nonumber\\
 \eql \int\dr^2\tht\;\frac{1}{2}\cW^2+\hc \nonumber\\
 &&+\int\dr^4\tht\;2\brc{2\brkt{\bar{\der}V-\bar{\Sgm}}\brkt{\der V-\Sgm}
 -\bar{\der}V\der V}, 
\eea
where we have used $d^2\bar{\tht}=-\frac{1}{4}\bar{D}^2$, 
and dropped total derivative terms. 
This agrees with (2.17) in Ref.~\cite{ArkaniHamed:2001tb} 
after rescaling the superfields as $V\to\frac{1}{\sqrt{2}g}V$ and 
$\Sgm\to\frac{1}{2g}\phi$.

\subsection{Dimensional reduction to 5D}
Here we consider the dimensional reduction of (\ref{expr:cL_VT2}) to five dimensions 
by neglecting the $x^5$-dependence of the $N=1$ superfields. 
Then (\ref{cstrt:Phi_T}) becomes 
\bea
 D^\alp\cW_{T\alp} \eql -2\der_4\Phi_T, \nonumber\\
 \bar{D}^2D^\alp\Phi_T \eql -4\der_4\cW_T^\alp.  
 \label{cstrt:Phi_T:5D}
\eea
Since the right-hand-side of the first equation is now real, 
$\cW_T^\alp$ satisfies the Bianchi 
identity~$D^\alp\cW_{T\alp}=\bar{D}_{\dalp}\bar{\cW}_T^{\dalp}$. 
Hence it is a field-strength superfield. 
\be
 \cW_T^\alp = -\frac{1}{4}\bar{D}^2D^\alp V_T, 
\ee
where $V_T$ is a real superfield. 
Substituting this into the second constraint in (\ref{cstrt:Phi_T:5D}), 
we obtain
\be
 \bar{D}^2D^\alp\brkt{\Phi_T-\der_4V_T} = 0, 
\ee
which indicates that
\be
 \Phi_T = \der_4 V_T-\Sgm_T-\bar{\Sgm}_T, \label{expr:Phi_T:5D}
\ee
where\footnote{
Note that $\Phi_T$ is a real linear superfield, \ie, $\Phi_T=\cPT\der_4V_T$. 
} 
\be
 \Sgm_T \equiv \frac{\bar{D}^2D^2}{16\Box_4}\der_4 V_T 
\ee
is a chiral part of $\der_4 V_T$. 
Then, the first constraint in (\ref{cstrt:Phi_T:5D}) is rewritten as
\be
 \brkt{\Box_4+\der_4^2}\cPT V_T = 0. 
\ee
Namely, the 6D tensor multiplet becomes an (on-shell) 5D vector multiplet 
after the dimensional reduction.\footnote{
Although there exists a 5D tensor field among the fields 
obtained from the 6D tensor field~$B_{MN}^+$ 
by the dimensional reduction, such a field is dual to 
a 5D vector field.  
The duality between the 5D tensor (gauge) multiplet and the 5D vector multiplet 
is explicitly shown in component fields in Ref.~\cite{Kugo:2002vc}.  
}

As shown in Appendix~\ref{5Dreduction}, 
the Lagrangian~(\ref{expr:cL_VT2}) becomes the following expression 
after the dimensional reduction. 
\bea
 \cL_{\rm VT}^{(5D)} \eql -\int\dr^2\tht\;C_{IJK}\Sgm^I\cW^J\cW^K+\hc \nonumber\\
 &&+\int\dr^4\tht\;\frac{C_{IJK}}{3}
 \brc{\brkt{\der_4 V^ID^\alp V^J-V^I\der_4D^\alp V^J}\cW_\alp^K+\hc} \nonumber\\
 &&+\int\dr^4\tht\;\frac{2C_{IJK}}{3}\cV^I\cV^J\cV^K, 
 \label{5DcL}
\eea
where $(\Sgm^1,V^1,\Sgm^2,V^2)=(\Sgm,V,\Sgm_T,V_T)$, 
the symmetric constant tensor~$C_{IJK}$ is defined as
$C_{112}=C_{121}=C_{211}=1$ and the other components are zero, and 
\be
 \cV^I \equiv \der_4 V^I-\Sgm^I-\bar{\Sgm}^I. 
\ee
This agrees with the 5D supersymmetric Chern-Simons 
terms~\cite{ArkaniHamed:2001tb,Hebecker:2008rk}.

\subsection{Bilinear terms in tensor multiplets}
In this subsection, we consider the Lagrangian terms that consist of 
only tensor multiplets. 
It is given by (\ref{L_tsr}). 
The Lagrangian is expressed as
\be
 \cL_{\rm tensor} = \oint_C\frac{d\zeta}{2\pi i\zeta}
 \int\dr^4\tht\;L_{\rm tensor} 
 = \int\dr^4\tht\;\cT_0^{[2]}|, \label{cL_tsr}
\ee
where
\be
 \cT^{[2]} = \zeta\cT_1^{[2]}+\cT_0^{[2]}-\frac{1}{\zeta}(\cT_1^{[2]})^*. 
\ee
Using the expressions in Sec.~\ref{N1tsr}, 
$\cT_0^{[2]}|$ is calculated as
\bea
 \cT_0^{[2]}| \eql \left.\brc{-i(D_\alp\Phi) T^{2\alp}
 +i(D'_\alp\Phi) T^{1\alp}+\hc}\right|
 +\frac{1}{4}\Phi_T\tl{\cA}_0 \nonumber\\
 \eql \brkt{-D^\alp\Phi_T\bar{D}^2\tl{Y}_\alp
 -\cW_T^\alp D_\alp\tl{X}+\hc}+\frac{1}{4}\Phi_T\tl{\cA}_0, 
\eea
where
\be
 \tl{\cA}_0 = -8\der\bar{\tl{X}}-4\bar{D}_{\dalp}D^2\bar{\tl{Y}}^{\dalp}+\tl{\phi}_0. 
\ee
Here we treat two tensor multiplets~$(\Phi_T,\cW_T^\alp)$ originating from $\Phi$ 
and $(\tl{X},\tl{Y}^\alp)$ originating from $T^{i\ualp}$ 
as independent multiplets. 
The Lagrangian~(\ref{cL_tsr}) is then expressed as
\bea
 \cL_{\rm tensor} \eql \int\dr^4\tht\;
 \brc{\brkt{\Phi_TD^\alp\bar{D}^2\tl{Y}_\alp+D^\alp\cW_{T\alp}\tl{X}+\hc}
 +\frac{1}{4}\Phi_T\tl{\cA}_0} \nonumber\\
 \eql \int\dr^4\tht\;\brkt{\frac{1}{8}\Phi_T\tl{\cA}_0
 +\Phi_TD^\alp\bar{D}^2\tl{Y}_\alp
 -2\bar{\der}\Phi_T\tl{X}+\hc} \nonumber\\
 \eql \int\dr^4\tht\;\frac{1}{8}\Phi_T\brkt{
 \tl{\cA}_0+8D^\alp\bar{D}^2\tl{Y}_\alp+16\bar{\der}\tl{X}+\hc} \nonumber\\
 \eql \int\dr^4\tht\;\Phi_T\brkt{\frac{1}{2}D^\alp\bar{D}^2\tl{Y}_\alp
 +\bar{\der}\tl{X}+\hc}.  
\eea
We have used (\ref{cstrt:Phi_T}), 
and dropped total derivative terms. 
At the last step, we have used that
\be
 \int\dr^4\tht\;\Phi_T\brkt{\tl{\cA}_0+4\bar{D}_{\dalp}D^2\bar{\tl{Y}}^{\dalp}
 +8\der\bar{\tl{X}}} 
 = \int\dr^4\tht\;\Phi_T\tl{\phi}_0 = 0. 
\ee
This Lagrangian can be further rewritten as
\bea
 \cL_{\rm tensor} \eql \int\dr^4\tht\;
 \brkt{-\frac{1}{2}\bar{D}^2D^\alp\Phi_T\tl{Y}_\alp
 -\bar{\der}\Phi_T\tl{X}+\hc} \nonumber\\
 \eql \int\dr^4\tht\;
 \brkt{-\frac{1}{2}\bar{D}^2D^\alp\Phi_T\tl{Y}_\alp
 -\frac{1}{2}\cW_T^\alp D_\alp\tl{X}+\hc} \nonumber\\
 \eql -\frac{1}{2}\int\dr^4\tht\;
 \brc{\bar{D}^2D^\alp\Phi_T\tl{Y}_\alp+\cW_T^\alp
 \brkt{\tl{Z}_\alp-4\der\tl{Y}_\alp}+\hc} \nonumber\\
 \eql -\frac{1}{2}\int\dr^4\tht\;\brc{
 \brkt{\bar{D}^2D^\alp\Phi_T+4\der\cW_T^\alp}\tl{Y}_\alp+\hc}. 
 \label{cL_tsr2}
\eea
At the last step, we have used the fact that $\cW_T^\alp$ and $\tl{Z}^\alp$ are 
a chiral and a linear superfields. 
This Lagrangian vanishes due to the second constraint in (\ref{cstrt:Phi_T}). 
However we can relax that constraint 
if we regard $\tl{Y}_\alp$ as the Lagrange multiplier. 
%(\ref{cL_tsr2}) is added to the Lagrangian. 
In that case, the constraint is obtained as the equation of motion for $\tl{Y}_\alp$. 
As shown in Appendix~\ref{cstrt:tsr}, that constraint is necessary 
in order for $\cF^{[2]}$ defined in (\ref{def:cF2}) to satisfy 
$\cD_\alp^{[1]}\cF^{[2]}=0$, which is relevant 
to the $N=2$ SUSY invariance of the action. 
Thus, in such a case, the full $N=2$ SUSY invariance of 
the vector-tensor coupling terms~(\ref{expr:cL_VT2}) 
is ensured only at the on-shell level.\footnote{
Half of the whole SUSY remains manifest at the off-shell level 
because the action is expressed in terms of 
$N=1$ superfields. 
}
Nevertheless, (\ref{cL_tsr2}) is expected to play an important role 
when we promote the theory to SUGRA. 
It corresponds to (2.14) in Ref.~\cite{Sokatchev:1988aa}, 
which is described in the harmonic superspace.

\subsection{Identification of component fields}
Finally, we identify component fields of each $N=1$ superfield. 
Here we focus on the bosonic fields. 

A 6D vector field~$A_M$ is embedded into $V$ and $\Sgm$ as~\cite{ArkaniHamed:2001tb}
\bea
 V \eql -(\tht\sgm^\mu\bar{\tht})A_\mu+\cdots, \nonumber\\
 \Sgm \eql \frac{1}{2}\brkt{A_5-iA_4}+\cdots, 
\eea
where the ellipses denote fermionic or auxiliary fields. 
\ignore{
Thus we obtain 
\bea
 \Sgm\cW^\alp \eql \frac{1}{2}(\tht\sgm^{\mu\nu})^\alp\brkt{A_4+iA_5}F_{\mu\nu}
 +\cdots,  \nonumber\\
 D^\alp V\cW_\alp \eql -i(\bar{\tht}\bar{\sgm}^\mu\sgm^{\nu\rho}\tht)A_\mu F_{\nu\rho}
 -\frac{1}{4}\tht^2\bar{\tht}^2
 \brkt{F_{\mu\nu}F^{\mu\nu}+\frac{i}{2}\ep^{\mu\nu\rho\sgm}F_{\mu\nu}F_{\rho\sgm}}
 +\cdots, \nonumber\\
 \der V D^\alp V \eql \frac{1}{2}\bar{\tht}^2(\tht\sgm^\mu\bar{\sgm}^\nu)^\alp
 \der A_\mu A_\nu+\cdots,  
\eea
where $F_{\mu\nu}\equiv\der_\mu A_\nu-\der_\nu A_\mu$ 
and $\sgm^{\mu\nu}\equiv\frac{1}{4}(\sgm^\mu\bar{\sgm}^\nu-\sgm^\nu\bar{\sgm}^\mu)$. 
}

The 6D tensor multiplet contains a real scalar field~$\sgm$ and 
a self-dual tensor field~$B_{MN}^+$, which satisfy~\cite{Sokatchev:1988aa} 
\bea
 \brkt{\Box_4+\der\bar{\der}}\sgm \eql 0, \nonumber\\
 \der_{[M}B_{NL]}^+ \eql \frac{1}{6}\ep_{MNLPQR}\der^P B^{+QR}, 
\eea
where $\ep_{MNLPQR}$ is the antisymmetric constant tensor. 
From (\ref{expr:T^i2}) and (\ref{expr:F^i}), 
the gauge transformation for the tensor multiplet in (\ref{gauge_trf}) 
is expressed as
\bea
 \dlt_G\brkt{Z_\alp-4\der Y_\alp} \eql D_\alp\brkt{\der V_G-\Sgm_G}, 
 \nonumber\\
 \dlt_G\bar{D}^2Y_\alp \eql -\frac{1}{4}\bar{D}^2D_\alp V_G, 
\eea
where the vector multiplet~$(\Sgm_G,V_G)$ is the transformation parameter 
that satisfies the on-shell condition. 
Note that $\Phi_T$ and $\cW_{T\alp}$ are invariant under this transformation. 
In components, this gauge transformation is expressed as 
\be
 \sgm \to \sgm, \;\;\;\;\;
 B^+_{MN} \to B^+_{MN}+\der_M\lmd_N-\der_N\lmd_M,  \label{gauge_trf:comp}
\ee
where $\lmd_M$ are the transformation parameters. 
From the conditions~$D^2\Phi_T=\bar{D}^2\Phi_T=0$ and (\ref{cstrt:Phi_T}), 
we find that 
$\sgm$ and $B_{MN}^+$ are embedded into $\Phi_T$ and $\cW_{T\alp}$ as
\bea
 \Phi_T \eql \sgm-2(\tht\sgm^\mu\bar{\tht})\brc{
 \der_\mu B_{45}^+-\Im(\der C_\mu)}
 -\frac{1}{4}\tht^2\bar{\tht}^2\Box_4\sgm+\cdots, \nonumber\\
 \cW_{T\alp} \eql \tht_\alp\bar{\der}\sgm
 +(\sgm^{\mu\nu}\tht)_\alp\brc{
 \bar{\der}B_{\mu\nu}^++\der_\mu C_\nu-\der_\nu C_\mu}+\cdots, 
\eea
where $C_\mu\equiv B_{\mu 4}^++iB_{\mu 5}^+$,  
and $\cW_{T\alp}$ is expressed 
in the chiral basis~$(x^\mu+i\tht\sgm^\mu\bar{\tht},z,\bar{z},\tht,\bar{\tht})$. 
Note that these expressions are invariant under (\ref{gauge_trf:comp}).

\section{Summary} \label{summary}
We have derived the $N=1$ superfield description of 
supersymmetric coupling terms among 6D tensor and vector multiplets 
from the projective superspace action provided in Ref.~\cite{Linch:2012zh}. 
This is necessary to describe 6D SUGRA in terms of $N=1$ superfields. 
Our result contains the result in Ref.~\cite{ArkaniHamed:2001tb} 
as a special case. 
It also reproduces the 5D supersymmetric Chern-Simons terms 
after the dimensional reduction. 

The tensor multiplet is described by two complex spinor 
superfields~$Y_\alp$ and $Z_\alp$, where $Z_\alp$ is constrained 
as $\bar{D}^2Z_\alp=0$. 
They appear in the action in the forms of a real linear superfield~$\Phi_T$ 
and a chiral spinor superfield~$\cW_{T\alp}$ defined by (\ref{def:PhiT-cWT}). 
These superfields are constrained by (\ref{cstrt:Phi_T}), 
which leads to the on-shell conditions. 
Thus they should be treated as external fields. 
This stems from the fact that the 6D tensor multiplet contains 
a self-dual tensor field~$B^+_{MN}$. 
As shown in Ref.~\cite{Bergshoeff:1985mz} in the component fields, 
the on-shell condition for the tensor multiplet can be relaxed 
when the theory couples to the gravity. 
Our result~(\ref{expr:cL_VT2}) provides a good starting point 
to obtain the $N=1$ superfield description of 6D SUGRA. 
We will discuss this issue in the subsequent paper.

\subsection*{Acknowledgements}
H.A., Y.S., and Y.Y. are supported in part by Grant-in-Aid for Young Scientists (B) 
(No. 25800158),
Grant-in-Aid for Scientific Research (C) (No.25400283),    
and Research Fellowships for Young Scientists (No.26-4236), 
which are from Japan Society for the Promotion of Science, respectively.  

\appendix

\ignore{
\section{Notations}
\subsection{Gamma matrices and conjugation matrices}
The spacetime metric is 
\be
 ds^2 = \eta_{MN}dx^Mdx^N = \eta_{\mu\nu}dx^\mu dx^\nu+(dx^4)^2+(dx^5)^2,  
\ee
where $\eta_{\mu\nu}=\diag(-1,1,1,1)$. 
The 6D gamma matrices~$\Gm^M$ ($M=0,1,\cdots,5$) are chosen as
\be
 \Gm^M = \begin{pmatrix} & \gm^M \\ \gm^M & \end{pmatrix} \;\;\;\;\; 
 (M \neq 5), \;\;\;\;\;
 \Gm^5 = \begin{pmatrix}  & \id_4 \\ -\id_4 &  \end{pmatrix}, 
\ee
where~\footnote{
We follow the notations of Ref.~\cite{Wess:1992cp} 
for 2-component spinor indices. 
} 
\be
 (\gm^\mu)_{\ualp}^{\;\:\ubt} = \begin{pmatrix} & \sgm^\mu_{\alp\dbt} \\
 \bar{\sgm}^{\mu\dalp\bt} & \end{pmatrix}, \;\;\;\;\;
 (\gm^4)_{\ualp}^{\;\;\ubt} = \begin{pmatrix} -i\dlt_\alp^{\;\;\bt} & \\
 & i\dlt^{\dalp}_{\;\;\dbt} \end{pmatrix}. 
\ee
These satisfy 
\be
 \brc{\Gm^M,\Gm^N} = -2\eta^{MN}. 
\ee
The 6D chirality matrix~$\Gm_7$ is defined as
\be
 \Gm_7 = \Gm^0\Gm^1\cdots\Gm^5 = \begin{pmatrix} \id_4 & \\ & -\id_4 
 \end{pmatrix}. 
\ee
The charge conjugation matrix~$\hat{C}$ is
\be
 \hat{C} = \begin{pmatrix} & -C \\ C & 0 \end{pmatrix}, \;\;\;\;\;
 C^{\ualp\ubt} = \begin{pmatrix} \ep^{\alp\bt} & \\ & -\ep_{\dalp\dbt} \end{pmatrix}, 
 \;\;\;\;\;
 (C^{-1})_{\ualp\ubt} = \begin{pmatrix} \ep_{\alp\bt} & \\ & -\ep^{\dalp\dbt} 
 \end{pmatrix}, 
\ee
where $\ep^{12}=\ep_{21}=1$. 
Then, it follows that
\be
 \hat{C}\Gm^M\hat{C}^{-1} = -(\Gm^M)^t, \;\;\;\;\;
 C\gm^MC^{-1} = (\gm^M)^t. 
\ee
Note that the charge conjugation flips the 6D chirality. 
The gamma matrices~$\Gm_{\rm LT}^M$ in Ref.~\cite{Linch:2012zh} are related to ours through
\be
 \Gm_{\rm LT}^M = \begin{pmatrix} \id_4 & \\ & C \end{pmatrix}\Gm^M
 \begin{pmatrix} \id_4 & \\ & C^{-1} \end{pmatrix}. 
\ee
Here we define $\hat{B}$ as 
\bea
 \hat{B} \defa \hat{C}\Gm^0 = \begin{pmatrix} B & \\ & -B \end{pmatrix}, \nonumber\\
 B_{\ualp}^{\;\;\bar{\bt}} \defa \begin{pmatrix} & \ep_{\alp\bt} \\ 
 -\ep^{\dalp\dbt} & \end{pmatrix}, \;\;\;\;\;
 (B^*)_{\bar{\alp}}^{\;\;\ubt} = \begin{pmatrix} & \ep_{\dalp\dbt} \\
 -\ep^{\alp\bt} & \end{pmatrix}, 
\eea
which satisfy 
\bea
 \hat{B}\hat{B}^* \eql \hat{B}^*\hat{B} = -\id_8, \;\;\;\;\;
 BB^* = B^*B = -\id_4, \nonumber\\
 \hat{B}^*\Gm^M\hat{B} \eql (\Gm^M)^*, \;\;\;\;\;
 B^*\gm^M B = -(\gm^M)^*. 
\eea
The indices~$\bar{\alp},\bar{\bt},\cdots$ denote those of 
the complex conjugate of 6D spinors. 
By using this matrix, the covariant conjugate of 
a 6D spinor~$\hat{\Psi}$ is defined as
\be
 \overline{\hat{\Psi}} \equiv \hat{B}\hat{\Psi}^*. 
\ee
Note that this operation preserves the 6D chirality, 
but it is not a $Z_2$ transformation since 
\be
 \overline{\overline{\hat{\Psi}}} = \overline{\hat{B}\hat{\Psi}^*}
 = \hat{B}\hat{B}^*\hat{\Psi} = -\hat{\Psi}. 
\ee
For an SU(2)-doublet spinor~$\hat{\Psi}^i$ ($i=1,2$), 
a $Z_2$ transformation is obtained by combining the covariant conjugation 
with lowering the SU(2) index, 
\be
 \hat{\Psi}^i \to \ep^{ij}\overline{\hat{\Psi}^j} 
 = \ep^{ij}\hat{B}(\hat{\Psi}^j)^*. 
\ee
Thus we can impose the SU(2)-Majorana condition, 
\be
 \ep^{ij}\overline{\hat{\Psi}^j} = \hat{\Psi}^i \;\;\;\;\;
 \Leftrightarrow \;\;\;\;\;
 \overline{\hat{\Psi}^i} = \hat{\Psi}_i \equiv \ep_{ij}\hat{\Psi}^j. 
\ee
Since the covariant conjugation preserves the 6D chirality, we can impose 
this condition on 6D Weyl spinors, 
\be
 \hat{\Psi}^i_+ = \begin{pmatrix} \Psi^i_+ \\ 0 \end{pmatrix}, \;\;\;\;\;
 \hat{\Psi}^i_- = \begin{pmatrix} 0 \\ i\Psi^i_- \end{pmatrix}, 
\ee
where the signs denote the 6D chiralities. 
Namely, the SU(2)-Majorana-Weyl condition is expressed as
\be
 \overline{\Psi^i_\pm} \equiv B(\Psi_\pm^i)^* = \Psi_{\pm i} \equiv \ep_{ij}\Psi_\pm^j. 
 \label{4comp_SU2MW}
\ee
A 4-component spinor~$\Psi^i_{\ualp}$ that satisfies this condition is written as
\be
 \Psi_{\ualp}^1 = \begin{pmatrix} \chi_\alp \\ \bar{\lmd}^{\dalp} \end{pmatrix}, \;\;\;\;\;
 \Psi_{\ualp}^2 = \begin{pmatrix} -\lmd_\alp \\ \bar{\chi}^{\dalp} \end{pmatrix}. 
\ee
\subsection{Spinor derivatives}
We introduce the Grassmann coordinates~$\Tht_{\ualp}^i$, which is 
an SU(2)-Majorana-Weyl condition with the 6D chirality~$-$.  
Then the covariant spinor derivatives are defined as
\be
 \cD_{\ualp}^i \equiv \ep^{ij}(C^{-1})_{\ualp\ubt}\frac{\der}{\der\Tht_{\ubt}^j}
 +i(\gm^M)_{\ualp}^{\;\;\ubt}\Tht^i_{\ubt}\der_M, 
 \label{def:cD}
\ee
which satisfies
\be
 \brc{\cD_{\ualp}^i,\cD_{\ubt}^j} = 
 2i\ep^{ij}(\gm^M C^{-1})_{\ualp\ubt}\der_M.  \label{cDcD}
\ee
In the 2-component spinor notation, $\Tht_{\ualp}^i$ are decomposed as
\be
 \Tht_{\ualp}^1 = \begin{pmatrix} \tht'_\alp \\ \bar{\tht}^{\dalp} \end{pmatrix}, 
 \;\;\;\;\;
 \Tht_{\ualp}^2 = \begin{pmatrix} -\tht_\alp \\ \bar{\tht}^{\prime\dalp} \end{pmatrix}. 
\ee
Then, the covariant spinor derivatives are expressed as
\be
 \cD_{\ualp}^1 = \begin{pmatrix} D_\alp \\ \bar{D}'^{\dalp} \end{pmatrix}, \;\;\;\;\;
 \cD_{\ualp}^2 = \begin{pmatrix} D'_\alp \\ -\bar{D}^{\dalp} \end{pmatrix}, 
 \label{decomp:cD}
\ee
where
\bea
 D_\alp \eql \frac{\der}{\der\tht^\alp}+i\brkt{\sgm^\mu\bar{\tht}}_\alp\der_\mu
 +\tht'_\alp\bar{\der}, \nonumber\\
 D'_\alp \eql \frac{\der}{\der\tht^{\prime\alp}}
 +i\brkt{\sgm^\mu\bar{\tht}'}_\alp\der_\mu-\tht_\alp\bar{\der}, \nonumber\\
 \bar{D}^{\dalp} \eql \frac{\der}{\der\bar{\tht}_{\dalp}}
 +i\brkt{\bar{\sgm}^\mu\tht}^{\dalp}\der_\mu+\bar{\tht}^{\prime\dalp}\der, 
 \nonumber\\
 \bar{D}^{\prime\dalp} \eql \frac{\der}{\der\bar{\tht}'_{\dalp}}
 +i\brkt{\bar{\sgm}^\mu\tht'}^{\dalp}\der_\mu
 -\bar{\tht}^{\dalp}\der, 
 \label{def:DbD}
\eea
and $\der\equiv\der_4-i\der_5$. 
The algebra~(\ref{cDcD}) is decomposed as
\bea
 \brc{D_\alp,\bar{D}_{\dalp}} \eql -2i\sgm^\mu_{\alp\dalp}\der_\mu, \;\;\;\;\;
 \brc{D'_\alp,\bar{D}'_{\dalp}} = -2i\sgm^\mu_{\alp\dalp}\der_\mu, \nonumber\\
 \brc{D_\alp,D'_\bt} \eql 2\ep_{\alp\bt}\bar{\der}, \;\;\;\;\;
 \brc{\bar{D}^{\dalp},\bar{D}^{\prime\dbt}} = 2\ep^{\dalp\dbt}\der, \nonumber\\
 \brc{D_\alp,D_\bt} \eql \brc{D'_\alp,D'_\bt} 
 = \brc{D'_\alp,\bar{D}_{\dalp}} = 0.  \label{Dalgebra}
\eea
}

\section{Notations for spinors} \label{notations}
\subsection{Gamma matrices}
The spacetime metric is 
\be
 ds^2 = \eta_{MN}dx^Mdx^N = \eta_{\mu\nu}dx^\mu dx^\nu+(dx^4)^2+(dx^5)^2,  
\ee
where $\eta_{\mu\nu}=\diag(-1,1,1,1)$. 

The 6D gamma matrices~$\Gm^M$ ($M=0,1,\cdots,5$) are chosen as
\be
 \Gm^M = \begin{pmatrix} & (\gm^M)_{\ualp\ubt} \\ (\tl{\gm}^M)^{\ualp\ubt} & 
 \end{pmatrix}, 
\ee
where $4\times 4$ matrices~$\gm^M$ and $\tl{\gm}^M$ satisfy
\bea
 \brkt{\gm^M\tl{\gm}^N+\gm^N\tl{\gm}^M}_{\ualp}^{\;\;\ubt} 
 \eql -2\eta^{MN}\dlt_{\ualp}^{\;\;\ubt}, \nonumber\\
 \brkt{\tl{\gm}^M\gm^N+\tl{\gm}^N\gm^M}^{\ualp}_{\;\;\ubt}
 \eql -2\eta^{MN}\dlt^{\ualp}_{\;\;\ubt}, 
\eea
and are defined as 
%\footnote{
%The definition of $\gm^5$ and $\tl{\gm}^5$ is different 
%from that of Ref.~\cite{Linch:2012zh} by a factor~$-1$. }
\bea
 (\gm^\mu)_{\ualp\ubt} \eql \begin{pmatrix} & -\sgm^\mu_{\alp\dgm}\ep^{\dgm\dbt} \\
 \bar{\sgm}^{\mu\dalp\gm}\ep_{\gm\bt} & \end{pmatrix}, \nonumber\\
 (\gm^4)_{\ualp\ubt} \eql \begin{pmatrix} -i\ep_{\alp\bt} & \\
 & -i\ep^{\dalp\dbt} \end{pmatrix}, \;\;\;\;\;
 (\gm^5)_{\ualp\ubt} = \begin{pmatrix} \ep_{\alp\bt} & \\ & -\ep^{\dalp\dbt} 
 \end{pmatrix}, \nonumber\\
 (\tl{\gm}^\mu)^{\ualp\ubt} \eql \begin{pmatrix} & \ep^{\alp\gm}\sgm^\mu_{\gm\dbt} \\
 -\ep_{\dalp\dgm}\bar{\sgm}^{\mu\dgm\bt} & \end{pmatrix}, \nonumber\\
 (\tl{\gm}^4)^{\ualp\ubt} \eql \begin{pmatrix} -i\ep^{\alp\bt} & \\
 & -i\ep_{\dalp\dbt} \end{pmatrix}, \;\;\;\;\;
 (\tl{\gm}^5)^{\ualp\ubt} = \begin{pmatrix} -\ep^{\alp\bt} & \\
 & \ep_{\dalp\dbt} \end{pmatrix}, 
\eea
where the antisymmetric tensors~$\ep^{\alp\bt}$ and $\ep_{\alp\bt}$ 
are chosen as $\ep^{12}=\ep_{21}=1$. 
These matrices are anti-symmetric, \ie, $(\gm^M)_{\ualp\ubt}=-(\gm^M)_{\ubt\ualp}$ 
and $(\tl{\gm}^M)^{\ualp\ubt}=-(\tl{\gm}^M)^{\ubt\ualp}$. 
The 6D chirality matrix~$\Gm_7$ is defined by 
\be
 \Gm_7 \equiv \Gm^0\Gm^1\Gm^2\Gm^3\Gm^4\Gm^5 = \begin{pmatrix} \id_4 & \\ & -\id_4 
 \end{pmatrix}. 
\ee

The antisymmetric tensors~$\ep_{\ualp\ubt\ugm\udlt}$ 
and $\ep^{\ualp\ubt\ugm\udlt}$ are given by
\be
 \ep_{\ualp\ubt\ugm\udlt} = \frac{1}{2}(\gm^M)_{\ualp\ubt}(\gm_M)_{\ugm\udlt}, 
 \;\;\;\;\;
 \ep^{\ualp\ubt\ugm\dlt} = \frac{1}{2}(\tl{\gm}^M)^{\ualp\ubt}(\tl{\gm}_M)^{\ugm\udlt}. 
\ee
Then it follows that
\be
 \frac{1}{2}\ep^{\ualp\ubt\ugm\udlt}(\gm^M)_{\ugm\udlt} = (\tl{\gm}^M)^{\ualp\ubt}, 
 \;\;\;\;\;
 \frac{1}{2}\ep_{\ualp\ubt\ugm\udlt}(\tl{\gm}^M)^{\ugm\udlt} 
 = (\gm^M)_{\ualp\ubt}. 
\ee
Since $\ep^{\underline{1}\underline{2}\underline{3}\underline{4}}
=1=-\ep^{12}\ep_{\dot{1}\dot{2}}$ 
and $\ep_{\underline{1}\underline{2}\underline{3}\underline{4}}
=1=-\ep_{12}\ep^{\dot{1}\dot{2}}$, 
these tensors are expressed in the 2-component notation as
\bea
 \ep^{\ualp\ubt\ugm\udlt} \eql -\ep^{\alp\bt}\ep_{\dgm\ddlt}
 -\ep^{\alp\gm}\ep_{\ddlt\dbt}-\ep^{\alp\dlt}\ep_{\dbt\dgm}
 -\ep_{\dalp\dbt}\ep^{\gm\dlt}-\ep_{\dalp\dgm}\ep^{\dlt\bt}
 -\ep_{\dalp\ddlt}\ep^{\bt\gm}, \nonumber\\
 \ep_{\ualp\ubt\ugm\udlt} \eql -\ep_{\alp\bt}\ep^{\dgm\ddlt}
 -\ep_{\alp\gm}\ep^{\ddlt\dbt}-\ep_{\alp\dlt}\ep^{\dbt\dgm}
 -\ep^{\dalp\dbt}\ep_{\gm\dlt}-\ep^{\dalp\dgm}\ep_{\dlt\bt}
 -\ep^{\dalp\ddlt}\ep_{\bt\gm}. 
 \label{decomp:ep}
\eea

\subsection{Conjugation matrices}
An 8-component Dirac spinor~$\hat{\Psi}$ is decomposed 
into 4-component Weyl spinors as
\be
 \hat{\Psi} = \begin{pmatrix} \Psi^{(+)}_{\ualp} \\ \Psi^{(-)\ualp} 
 \end{pmatrix}, 
\ee
where the signs denote eigenvalues of $\Gm_7$. 
The Dirac conjugate of $\hat{\Psi}$ is defined as
\be
 \bar{\hat{\Psi}} \equiv \hat{\Psi}^\dagger A, 
\ee
where $\hat{A}$ satisfies 
\be
 \hat{A}\Gm^M\hat{A}^{-1} = (\Gm^M)^\dagger. 
\ee
The explicit form of $\hat{A}$ is given by
\bea
 \hat{A} \eql \begin{pmatrix} & A \\ \tl{A} & \end{pmatrix}, \nonumber\\
 A^{\bar{\alp}}_{\;\;\ubt} \eql \begin{pmatrix} & \ep^{\dalp\dbt} \\
 -\ep_{\alp\bt} & \end{pmatrix}, \;\;\;\;\;
 \tl{A}_{\bar{\alp}}^{\;\;\ubt} = \begin{pmatrix} & \ep_{\dalp\dbt} \\
 -\ep^{\alp\bt} & \end{pmatrix}, 
\eea
where $\bar{\alp}$ denotes a 4-component spinor index of 
the complex conjugate of the Weyl spinors. 

Since $(\Gm^M)^*$ form an equivalent representation of the Clifford algebra, 
there exists an invertible matrix~$\hat{B}$ that satisfies  
\be
 \hat{B}(\Gm^M)^*\hat{B}^t = \Gm^M. 
\ee
An explicit form of $\hat{B}$ is given by
\be
 \hat{B} = \begin{pmatrix} B & \\ & B^\dagger \end{pmatrix}, 
\ee
where 
\be
 B_{\ualp}^{\;\;\bar{\bt}} \equiv \begin{pmatrix} & \ep_{\alp\bt} \\
 -\ep^{\dalp\dbt} & \end{pmatrix}, \;\;\;\;\;
 (B^*)_{\bar{\alp}}^{\;\;\ubt} 
 = \begin{pmatrix} & \ep_{\dalp\dbt} \\ -\ep^{\alp\bt} & \end{pmatrix}. 
\ee
These matrices satisfy
\be
 BB^* = B^*B = -\id_4, \;\;\;\;\;
 B(\gm^M)^*B^* = \gm^M. 
\ee

The charge conjugation matrix~$\hat{C}$, which satisfies 
\be
 \hat{C}\Gm^M\hat{C}^{-1} = -(\Gm^M)^t, 
\ee
is constructed 
from $\hat{A}$ and $\hat{B}$ as
\be
 \hat{C} \equiv \hat{B}^\dagger\hat{A} 
 = \begin{pmatrix} & C \\ \tl{C} & \end{pmatrix},  
\ee
where 
\be
 C^{\ualp}_{\;\;\ubt} = \begin{pmatrix} -\dlt^\alp_{\;\;\bt} & \\
 & -\dlt_{\dalp}^{\;\;\dbt} \end{pmatrix}, \;\;\;\;\;
 \tl{C}_{\ualp}^{\;\;\ubt} = \begin{pmatrix} -\dlt_\alp^{\;\;\bt} & \\
 & -\dlt^{\dalp}_{\;\;\dbt} \end{pmatrix}. 
\ee
Thus the charge conjugation flips the 6D chirality. 

The covariant conjugate of a spinor~$\hat{\Psi}$ is defined as
\be
 \overline{\hat{\Psi}} \equiv \hat{B}\hat{\Psi}^*.  \label{def:ovln}
\ee
This operation is not a $Z_2$ transformation since
\be
 \overline{\overline{\hat{\Psi}}} = \overline{\hat{B}\hat{\Psi}^*}
 = \hat{B}(\hat{B}\hat{\Psi}^*)^* 
 = \hat{B}\hat{B}^*\hat{\Psi} = -\hat{\Psi}. 
\ee
For an SU(2)-doublet spinor~$\hat{\Psi}^i$ ($i=1,2$), 
a $Z_2$ transformation is obtained by combining the covariant conjugation 
with lowering the SU(2) index, 
\be
 \hat{\Psi}^i \to \ep^{ij}\overline{\hat{\Psi}^j} 
 = \ep^{ij}\hat{B}(\hat{\Psi}^j)^*. 
\ee
Thus we can impose the SU(2)-Majorana condition, 
\be
 \ep^{ij}\overline{\hat{\Psi}^j} = \hat{\Psi}^i \;\;\;\;\;
 \Leftrightarrow \;\;\;\;\;
 \overline{\hat{\Psi}^i} = \hat{\Psi}_i \equiv \ep_{ij}\hat{\Psi}^j. 
 \label{SU2MWcond}
\ee
Here the antisymmetric tensors~$\ep^{ij}$ and $\ep_{ij}$ are chosen as 
$\ep^{12}=\ep_{21}=1$. 
Since the covariant conjugation preserves the 6D chirality, 
we can impose this condition on 6D Weyl spinors. 
Namely, the SU(2)-Majorana-Weyl condition is expressed 
in the 4-component-spinor notation as
\bea
 \brkt{\overline{\Psi^{(+)i}}}_{\ualp} \defa B_{\ualp}^{\;\;\bar{\bt}}
 (\Psi^{(+)i*})_{\bar{\bt}} = \Psi^{(+)}_{i\ualp} 
 \equiv \ep_{ij}\Psi^{(+)j}_{\ualp}, \nonumber\\
 \brkt{\overline{\Psi^{(-)i}}}^{\ualp} \defa (B^\dagger)^{\ualp}_{\;\;\bar{\bt}}
 (\Psi^{(-)i*})^{\bar{\bt}} = \Psi^{(-)\ualp}_i
 \equiv \ep_{ij}\Psi^{(-)j\ualp}. 
 \label{4comp_SU2MW}
\eea
In the two-component-spinor notation, 
the SU(2)-Majorana-Weyl spinors are expressed as
\bea
 \Psi^{(+)1}_{\ualp} \eql \begin{pmatrix} \chi^{(+)}_\alp \\ 
 \bar{\lmd}^{(+)\dalp} \end{pmatrix}, \;\;\;\;\;
 \Psi^{(+)2}_{\ualp} = \begin{pmatrix} -\lmd^{(+)}_\alp \\
 \bar{\chi}^{(+)\dalp} \end{pmatrix}, \nonumber\\
 \Psi^{(-)1\ualp} \eql \begin{pmatrix} \chi^{(-)\alp} \\
 \bar{\lmd}^{(-)}_{\dalp} \end{pmatrix}, \;\;\;\;\;
 \Psi^{(-)2\ualp} = \begin{pmatrix} -\lmd^{(-)\alp} \\
 \bar{\chi}^{(-)}_{\dalp} \end{pmatrix}. 
\eea

\subsection{Covariant spinor derivatives}
We introduce the Grassmann coordinates~$\Tht^{i\ualp}$, 
which form an SU(2)-Majorana-Weyl spinor with the 6D chirality~$-$. 
Then the covariant spinor derivatives are defined as 
\be
 \cD_{\ualp}^i \equiv \ep^{ij}\frac{\der}{\der\Tht^{j\ualp}}
 +i(\gm^M)_{\ualp\ubt}\Tht^{i\ubt}\der_M 
 = -\frac{\der}{\der\Tht_i^{\ualp}}
 +i(\gm^M)_{\ualp\ubt}\Tht^{i\ubt}\der_M, 
 \label{def:cD}
\ee
which satisfies 
\be
 \brc{\cD^i_{\ualp},\cD^j_{\ubt}} = -2i\ep^{ij}
 (\gm^M)_{\ualp\ubt}\der_M.  \label{cDcD}
\ee
In the 2-component-spinor notation, $\Tht^{i\ualp}$ are 
expressed as
\be
 \Tht^{1\ualp} = \begin{pmatrix} \tht^{\prime\alp} \\ -\bar{\tht}_{\dalp}
 \end{pmatrix}, \;\;\;\;\;
 \Tht^{2\ualp} = \begin{pmatrix} \tht^\alp \\ \bar{\tht}'_{\dalp} 
 \end{pmatrix}.  \label{decomp:Tht}
\ee
Then, the covariant spinor derivatives are expressed as
\be
 \cD_{\ualp}^1 = \begin{pmatrix} D_\alp \\ \bar{D}'^{\dalp} \end{pmatrix}, \;\;\;\;\;
 \cD_{\ualp}^2 = \begin{pmatrix} -D'_\alp \\ \bar{D}^{\dalp} \end{pmatrix}, 
 \label{decomp:cD}
\ee
where
\bea
 D_\alp \eql \frac{\der}{\der\tht^\alp}+i\brkt{\sgm^\mu\bar{\tht}}_\alp\der_\mu
 +\tht'_\alp\bar{\der}, \nonumber\\
 D'_\alp \eql \frac{\der}{\der\tht^{\prime\alp}}
 +i\brkt{\sgm^\mu\bar{\tht}'}_\alp\der_\mu-\tht_\alp\bar{\der}, \nonumber\\
 \bar{D}^{\dalp} \eql \frac{\der}{\der\bar{\tht}_{\dalp}}
 +i\brkt{\bar{\sgm}^\mu\tht}^{\dalp}\der_\mu+\bar{\tht}^{\prime\dalp}\der, 
 \nonumber\\
 \bar{D}^{\prime\dalp} \eql \frac{\der}{\der\bar{\tht}'_{\dalp}}
 +i\brkt{\bar{\sgm}^\mu\tht'}^{\dalp}\der_\mu
 -\bar{\tht}^{\dalp}\der, 
 \label{def:DbD}
\eea
and $\der\equiv\der_4-i\der_5$. 
The algebra~(\ref{cDcD}) is decomposed as
\bea
 \brc{D_\alp,\bar{D}_{\dalp}} \eql -2i\sgm^\mu_{\alp\dalp}\der_\mu, \;\;\;\;\;
 \brc{D'_\alp,\bar{D}'_{\dalp}} = -2i\sgm^\mu_{\alp\dalp}\der_\mu, \nonumber\\
 \brc{D_\alp,D'_\bt} \eql 2\ep_{\alp\bt}\bar{\der}, \;\;\;\;\;
 \brc{\bar{D}^{\dalp},\bar{D}^{\prime\dbt}} = 2\ep^{\dalp\dbt}\der, \nonumber\\
 \brc{D_\alp,D_\bt} \eql \brc{D'_\alp,D'_\bt} 
 = \brc{D'_\alp,\bar{D}_{\dalp}} = 0.  \label{Dalgebra}
\eea
We list some useful formulae following from this algebra. 
\bea
 \sbk{D_\alp,\bar{D}^2} \eql -4i\sgm^\mu_{\alp\dalp}\der_\mu\bar{D}^{\dalp}, 
 \;\;\;\;\;
 \sbk{\bar{D}_{\dalp}D^2} = 4i\sgm^\mu_{\alp\dalp}\der_\mu D^\alp, \nonumber\\
 D'D \eql DD'-4\bar{\der}, \;\;\;\;\;
 \bar{D}'\bar{D} = \bar{D}\bar{D}'-4\der, \nonumber\\
 D_\alp D_\bt \eql \frac{1}{2}\ep_{\alp\bt}D^2, \;\;\;\;\;
 \bar{D}_{\dalp}\bar{D}_{\dbt} = -\frac{1}{2}\ep_{\dalp\dbt}\bar{D}^2. 
 \label{D:formulae}
\eea

\section{Constraints on $\bdm{\Phi_T}$ and $\bdm{\cW_T^\alp}$} \label{cstrt:tsr}
Here we derive the constraints in (\ref{cstrt:Phi_T}). 

From the definition~(\ref{def:PhiT-cWT}), 
\bea
 D^\alp\cW_{T\alp} \eql iD^\alp\bar{D}^2\brkt{D_\alp\bar{X}+4\bar{\der}Y_\alp} 
 \nonumber\\
 \eql i\bar{D}_{\dalp}D^2\bar{D}^{\dalp}\bar{X}+4i\bar{\der}D^\alp\bar{D}^2Y_\alp 
 \nonumber\\
 \eql i\bar{D}_{\dalp}D^2\brkt{\bar{Z}^{\dalp}-4\bar{\der}\bar{Y}^{\dalp}}
 +4i\bar{\der}D^\alp\bar{D}^2Y_\alp \nonumber\\
 \eql -2\bar{\der}\brkt{2i\bar{D}_{\dalp}D^2\bar{Y}^{\dalp}
 -2iD^\alp\bar{D}^2Y_\alp} = -2\bar{\der}\Phi_T. 
 \label{derive:DWT}
\eea
We have used that $\bar{D}^{\dalp}\bar{X}=\bar{Z}^{\dalp}-4\bar{\der}\bar{Y}^{\dalp}$ 
and $D^2\bar{Z}^{\dalp}=0$. 

The analyticity condition~$\cD_\alp^{[1]}\cF^{[2]}=0$ is 
translated in the 2-component-spinor notation as
\bea
 D_\alp\cF_1^{[2]} \eql 0, \;\;\;\;\;
 D_\alp\cF_0^{[2]}+D'_\alp\cF_1^{[2]} = 0, \nonumber\\
 \bar{D}'_{\dalp}\cF_1^{[2]} \eql 0, \;\;\;\;\;
 \bar{D}_{\dalp}\cF_1^{[2]}-\bar{D}'_{\dalp}\cF_0^{[2]} = 0. 
\eea
Thus the $N=1$ superfields~$\cF_0^{[2]}|$ and $\cF_1^{[2]}|$ satisfy 
the following constraints.  
\bea
 D_\alp\cF_1^{[2]}| \eql 0, \nonumber\\
 D^2\cF_0^{[2]}| \eql -DD'\cF_1^{[2]}| = -\brkt{D'D+4\bar{\der}}\cF_1^{[2]}| 
 = -4\bar{\der}\cF_1^{[2]}|. 
 \label{cstrt:cF_i^2}
\eea
From the explicit expressions in (\ref{expr:cF2i}),  
we can see that the first constraint is satisfied. 
As for the second constraint, we can show that
\be
 D^2\cF_0^{[2]}| = -4\bar{\der}\cF_1^{[2]}|
 -\brkt{D^2\bar{D}_{\dalp}\Phi_T+4\bar{\der}\bar{\cW}_{T\dalp}}\bar{\cW}^{\dalp}. 
\ee
We have used the constraint~(\ref{derive:DWT}). 
Comparing this with the second constraint in (\ref{cstrt:cF_i^2}), 
we obtain 
\be
 D^2\bar{D}_{\dalp}\Phi_T = -4\bar{\der}\bar{\cW}_{T\dalp}. 
\ee

\section{Derivation of 5D Lagrangian} \label{5Dreduction}
We derive (\ref{5DcL}) from (\ref{expr:cL_VT2}) 
after the dimensional reduction to 5D. 
By using (\ref{cstrt:Phi_T:5D}) and (\ref{expr:Phi_T:5D}), we can calculate 
\bea
 &&2\Phi_T\brc{V\brkt{\Box_4\cPT+\der_4^2}V
 +2\brkt{\der_4 V-\bar{\Sgm}}\brkt{\der_4 V-\Sgm}} \nonumber\\
% \eql 2\Phi_T\brc{\frac{1}{2}VD^\alp\cW_\alp
% +V\der_4^2V+2\brkt{\der_4 V-\bar{\Sgm}}\brkt{\der_4 V-\Sgm}} \nonumber\\
 \eql \brkt{\der_4 V_T-\Sgm_T-\bar{\Sgm}_T}VD^\alp\cW_\alp
 -2\brkt{\der_4\Phi_TV+\Phi_T\der_4V}\der_4V \nonumber\\
 &&+4\Phi_T\brc{(\der_4 V)^2-\der_4V\brkt{\Sgm+\bar{\Sgm}}+\bar{\Sgm}\Sgm} 
 \nonumber\\
 \eql \brc{\frac{1}{2}\der_4 V_TVD^\alp\cW_\alp
 -\bar{\Sgm}_TVD^\alp\cW_\alp
 +\frac{1}{2}D^\alp\cW_{T\alp}V\der_4V+\hc}
 \nonumber\\
 &&+2\Phi_T\brc{(\der_4V)^2-2\der_4V\brkt{\Sgm+\bar{\Sgm}}+2\bar{\Sgm}\Sgm} 
 \nonumber\\
 \eql \brc{-\frac{1}{2}D^\alp\brkt{\der_4 V_TV}\cW_\alp
 +\bar{\Sgm}_TD^\alp V\cW_\alp
 -\frac{1}{2}D^\alp\brkt{V\der_4V}\cW_{T\alp}+\hc}
 \nonumber\\
 &&+2\Phi_T\brkt{\der_4V-\Sgm-\bar{\Sgm}}^2-2\Phi_T\brkt{\Sgm^2+\bar{\Sgm}^2}. 
\eea
We have also used $D^\alp\cW_\alp=\bar{D}_{\dalp}\bar{\cW}^{\dalp}$, 
and dropped total derivative terms. 
Thus, after the dimensional reduction to 5D, (\ref{expr:cL_VT2}) becomes 
\bea
 \cL_{\rm VT}^{(5D)} \eql -\int\dr^2\tht\;\brc{2\Sgm\cW\cW_T
 +\frac{1}{4}\bar{D}^2\brkt{\Phi_TD^\alp V\cW_\alp
 +\der_4 VD^\alp V\cW_{T\alp}}}+\hc \nonumber\\
 &&+\int\dr^4\tht\;2\Phi_T\brc{V\brkt{\Box_4\cPT+\der_4^2}V
 +2\brkt{\der_4 V-\bar{\Sgm}}\brkt{\der_4 V-\Sgm}} \nonumber\\
 \eql -\int\dr^2\tht\;2\Sgm\cW\cW_T+\hc \nonumber\\
 &&+\int\dr^4\tht\;
 \brc{\brkt{\der_4V_T-\Sgm_T-\bar{\Sgm}_T}D^\alp V\cW_\alp
 +\der_4 VD^\alp V\cW_{T\alp}+\hc} \nonumber\\
 &&+\int\dr^4\tht\;\left\{
 \brkt{-\frac{1}{2}D^\alp\brkt{\der_4 V_TV}\cW_\alp
 +\bar{\Sgm}_TD^\alp V\cW_\alp
 -\frac{1}{2}D^\alp\brkt{V\der_4V}\cW_{T\alp}+\hc}
 \right.\nonumber\\
 &&\hspace{18mm}\left.
 +2\Phi_T\brkt{\der_4V-\Sgm-\bar{\Sgm}}^2-2\Phi_T\brkt{\Sgm^2+\bar{\Sgm}^2}
 \right\} \nonumber\\
 \eql -\int\dr^2\tht\;\brkt{2\Sgm\cW\cW_T+\Sgm_T\cW^2}+\hc \nonumber\\
 &&+\int\dr^4\tht\left\{
 \frac{1}{2}\brkt{\der_4V_T D^\alp V-\der_4 D^\alp V_TV}\cW_\alp 
 +\frac{1}{2}\brkt{\der_4VD^\alp V-\der_4 D^\alp VV}\cW_{T\alp}+\hc\right\} 
 \nonumber\\
 &&+\int\dr^4\tht\;2\brkt{\der_4V_T-\Sgm_T-\bar{\Sgm}_T}
 \brkt{\der_4 V-\Sgm-\bar{\Sgm}}^2. 
 \label{cL_VT^5D}
\eea 
We have dropped total derivative terms, 
and used that 
\be
 \int\dr^4\tht\;\Phi_T\Sgm^2=-\frac{1}{4}\int\dr^2\tht\;\bar{D}^2
 (\Phi_T\Sgm^2)=-\frac{1}{4}\int\dr^2\tht\;(\bar{D}^2\Phi_T)\Sgm^2=0. 
\ee
Using (\ref{Dalgebra}) and (\ref{D:formulae}), we can show that
\bea
 &&\brkt{\der_4V_TD^\alp V-\der_4D^\alp V_TV}\cW_\alp
 +\brkt{\der_4VD^\alp V-\der_4 D^\alp VV}\cW_{T\alp}+\hc \nonumber\\
 \eql 2\brkt{\der_4 VD^\alp V_T-\der_4D^\alp VV_T}\cW_\alp+\hc, 
\eea
up to total derivatives. 
Thus, (\ref{cL_VT^5D}) is rewritten as
\bea
 \cL_{\rm VT}^{(5D} \eql 
 -\int\dr^2\tht\;\brkt{2\Sgm\cW\cW_T+\Sgm_T\cW^2}+\hc \nonumber\\
 &&+\int\dr^4\tht\left\{
 \frac{1}{3}\brkt{\der_4V_T D^\alp V-\der_4 D^\alp V_TV}\cW_\alp 
 +\frac{1}{3}\brkt{\der VD^\alp V_T-\der_4 D^\alp VV_T}\cW_\alp \right.\nonumber\\
 &&\hspace{15mm}\left. 
 +\frac{1}{3}\brkt{\der_4VD^\alp V-\der_4 D^\alp VV}\cW_{T\alp}+\hc\right\} 
 \nonumber\\
 &&+\int\dr^4\tht\;2\brkt{\der_4V_T-\Sgm_T-\bar{\Sgm}_T}
 \brkt{\der_4 V-\Sgm-\bar{\Sgm}}^2. 
\eea
If we relabel $(\Sgm,V)$ and $(\Sgm_T,V_T)$ as $(\Sgm^1,V^1)$ and $(\Sgm^2,V^2)$, 
this is expressed as (\ref{5DcL}).

%%%%%%%%%%%%%%%%%%%%%%%%%%%% References %%%%%%%%%%%%%%%%%%%%%%%%%%%%%%

\end{document}